\documentclass[10pt,a4paper]{article}
\usepackage[english]{babel}
\usepackage[latin1]{inputenc}
\usepackage{amsfonts,amsbsy,bm,euscript,mathrsfs}
\usepackage{amssymb,stmaryrd,faktor}
\usepackage[tbtags]{amsmath}
\usepackage{bbm}
\usepackage{graphicx}
\usepackage[title,titletoc]{appendix}
\usepackage[bookmarks=true,colorlinks=true,linkcolor=blue,citecolor=blue,urlcolor=blue,bookmarksnumbered]{hyperref}

\usepackage{dsfont}
\usepackage{collref}
\usepackage{lmodern}
\usepackage{mathrsfs}
\usepackage{mathtools}
\usepackage{bbm}
\usepackage{braket}
\usepackage{slashed}

\usepackage{graphicx}
\usepackage{booktabs}
\usepackage{subfig}
\usepackage{tikz}
\usetikzlibrary{plotmarks,calc,decorations,decorations.pathmorphing}

\numberwithin{equation}{section}

\makeatletter
\renewcommand\section{\@startsection {section}{1}{\z@}
{-3.5ex \@plus -1ex \@minus -.2ex}
{2.3ex \@plus.2ex}
{\normalfont\Large\bfseries}}
\renewcommand\subsection{\@startsection{subsection}{2}{\z@}
{-3.25ex\@plus -1ex \@minus -.2ex}
{1.5ex \@plus.2ex}
{\normalfont\large\bfseries}}
\makeatother

\newcommand{\arXivlink}[1]{\href{http://arXiv.org/abs/#1}{arXiv:#1}}

\newcommand{\alg}[1]{\mathfrak{#1}}
\newcommand{\comm}[2]{[#1,#2]}
\newcommand{\acomm}[2]{\{#1,#2\}}

\def\e{\epsilon}

\usepackage{blkarray}

\begin{document}

\thispagestyle{empty}
\begin{flushright}\footnotesize\ttfamily
DMUS-MP-22/02
\end{flushright}
\vspace{2em}

\begin{center}

{\Large\bf \vspace{0.2cm}
{\color{black} \large On factorising twists in $AdS_3$ and $AdS_2$}} 
\vspace{1.5cm}

\textrm{Alessandro Torrielli\footnote{\texttt{a.torrielli@surrey.ac.uk}}}

\vspace{2em}

\vspace{1em}
\begingroup\itshape
Department of Mathematics, University of Surrey, Guildford, GU2 7XH, UK
\par\endgroup

\end{center}

\vspace{2em}

\begin{abstract}\noindent 
In this paper we study factorising twists of the massless $AdS_3$ and $AdS_2$ integrable $R$-matrices, and explore the programme of analysis of form factors which Maillet et al developed for ordinary spin-chains. We derive the factorising twists from the universal $R$-matrix of the $\alg{gl}(1|1)$ Yangian double, and discuss the RTT relations for the two- and three-site monodromy matrix. We show how the twist can be used to compute a simple scalar product. We then implement our construction in the language  of free fermions. Finally, we show how to obtain the massless $AdS_2$ quantum $R$-matrix from the Yangian universal $R$-matrix, and compute a peculiar factorising twist for this case as well.  

\end{abstract}

\newpage

\overfullrule=0pt
\parskip=2pt
\parindent=12pt
\headheight=0.0in \headsep=0.0in \topmargin=0.0in \oddsidemargin=0in

\vspace{-3cm}
\thispagestyle{empty}
\vspace{-1cm}

\tableofcontents

\setcounter{footnote}{0}

\section{Introduction}

\subsection{Integrability in $AdS_3$ and $AdS_2$ backgrounds}

\subsubsection{$AdS_3$}

Integrability in $AdS_3 \times S^3 \times M^4$ \cite{Bogdan}, with $M^4$ being either $T^4$ or $S^3 \times S^1$ \cite{Sundin:2012gc}, - see \cite{rev3,Borsato:2016hud} for reviews - has produced a vast number of results by walking in the footsteps of the originally established $AdS_5$ analysis \cite{Beisertreview,Foundations}. A vast array of articles has explored numerous aspects of this integrable system \cite{OhlssonSax:2011ms,seealso3,Borsato:2012ss,Borsato:2013qpa,Borsato:2013hoa,Rughoonauth:2012qd,PerLinus,CompleteT4,Borsato:2015mma,Beccaria:2012kb,Sundin:2013ypa,Bianchi:2013nra,Bianchi:2013nra1,Bianchi:2013nra2}. The undeniably new ingredient which features in $AdS_3$ are the massless excitations \cite{Sax:2012jv,Borsato:2016xns,Sax:2014mea,Baggio:2017kza} which call into the game the ideas behind the massless $S$-matrix and massless flows \cite{Zamol2,Fendley:1993jh,DiegoBogdanAle}. A substantial body of work has been produced in these investigations \cite{Lloyd:2013wza,Abbott:2012dd,Abbott:2014rca,MI,Abbott:2020jaa,Eberhardt:2017fsi,Gaber1,Gaber2,Gaber3,Gaber4,Gaber5,
GaberdielUltimo,Prin,Prin1,Abbott:2015mla,Per9,Hoare:2018jim,AntonioMartin,Regelskis:2015xxa,olobo,Baggio,JuanMiguelAle}). Recently, the Quantum Spectral Curve (QSC) has been adapted to the massive $AdS_3 \times S^3 \times T^4$ spectral problem in \cite{QSC} (see also references therein for an outline of the formalism). A new series of articles \cite{AleSSergey}, see also \cite{Seibold:2022mgg}, has revisited several aspects of $AdS_3 \times S^3 \times T^4$ $S$-matrix integrability and made novel proposals, most importantly for the dressing phases and notably the mirror TBA. The current paper will not be focusing on the dressing phases, and our results, merely pertaining to the Hopf algebra, will apply generally.  

The Hopf algebra structure typical of AdS/CFT \cite{Hopf}, when realised in the sector of massless excitations, very clearly shows the perspective of the quantised string theory as a $q$-deformation of the standard 2D Poincar\'e (super)symmetry coproduct. The articles \cite{CesarRafa,Charles} first explored this connection in the case of $AdS_5$, recently revisited in \cite{Riccardo}, see also \cite{qseealso}. When dealing with the case of $AdS_3$, this was put to fruition in \cite{JoakimAle,BorStromTorri,Andrea}. In \cite{gamma1,gamma2}, building upon these ideas, it was shown that there is a change of variables which manifests the difference-form of the massless $S$-matrix. This form is identical to the non-trivial BMN limit when taken for particles with the same worldsheet chirality \cite{DiegoBogdanAle}. The `pseudo-relativistic' variable suggested the massive version introduced in \cite{AleSSergey}. 

While the $q$-Poincar\'e deformation is just the ordinary string theory, re-written in a way which displays the built-in quantum group, in $AdS_3$ one can genuinely deform the model to introduce a mixture of RR and NSNS fluxes \cite{Cagnazzo:2012se,s1,s2,Babichenko:2014yaa,seealso12}. This has the effect of changing the single-particle energy $E$ vs. momentum $p$ dispersion relation, which becomes \cite{ArkadyBenStepanchuk,Lloyd:2014bsa}:
\begin{equation}
\label{eq:disp-rel}
E = \sqrt{\Big(m + \frac{k}{2 \pi} p\Big)^2 + 4 h^2 \sin^2\frac{p}{2}}.
\end{equation}
In this formula $k \in \mathbbmss{N}$ is the WZW level while $m$ is the particle-mass and $h$ is the coupling (see \cite{OhlssonSax:2018hgc} for a complete analysis of the string moduli space). In \cite{gamma2} a particular procedure on the massive mixed-flux $S$-matrix managed to produce another instance of massless relativistic $S$-matrices, describing critical theories whose central charge is encoded in a very compact set of TBA equations.

\subsubsection{$AdS_2$}

Integrability in $AdS_2 \times S^2 \times T^6$ \cite{ads2} has also been investigated in depth, in the attempt to catch a non-perturbative glimpse of the dual theory - either a superconformal quantum mechanics or a chiral CFT \cite{dual,gen}. Integrability has been demonstrated \cite{Sorokin:2011rr,Cagnazzo:2011at,amsw}, and in \cite{Hoare:2014kma} the putative $S$-matrix for massive and massless excitations was derived, based on a $\mathfrak{psu}(1|1)^2$ Hopf algebra with triple central extension. The (Yangian) representations which are relevant for massive magnons are long (typical) (cf. \cite{Arutyunov:2009pw}), while the massless modes form short representations \cite{Hoare:2014kma,Hoare:2014kmaa}. We should mention that both $AdS_3$ and $AdS_2$ still present some open questions in the matching with perturbation theory, see also \cite{Per6, Per10, MI,Borsato:2016xns}. 

The massive $S$-matrix is rather more complicated than its higher-dimensional counterparts, however in the massless limit it enormously simplifies. In \cite{Andrea2} a discussion was provided of how the BMN limit is distantly connected with ${\cal{N}}=1$ supersymmetric $S$-matrices \cite{Fendley:1990cy} via a particular limiting procedure. The $AdS_2$ $S$-matrix has the typical structure of 8-vertex models or of the XYZ spin-chain \cite{Baxter:1972hz, Baxter:1972hz1,Schoutens,MC} - in particular the algebraic Bethe ansatz is not directly applicable due to the lack of a reference state \cite{Levkovich-Maslyuk:2016kfv}, see also \cite{Nepo, Nepo1, Nepo2, Nepo3, Nepo4, Nepo5, Nepo6} (although a pseudovacuum is present for the tensor product of two copies). This problem was overcome in \cite{Andrea2}, see also the subsequent \cite{Ale}, by exploiting the so-called {\it free fermion} condition - as noticed in \cite{MC,Ahn} for these type of models - and by using the idea of inversion relations from \cite{Zamolodchikov:1991vh}. 

The free fermion condition holds in fact in far greater generality \cite{Lieb,MitevEtAl}. The article \cite{Dublin} afterwards produced a complete analysis and a comprehensive set of Bogoliubov transformations which recast the $AdS_3$ - and to a more limited extent the $AdS_2$ and $AdS_5$ - integrable systems in terms of free fermionic creation and annihilation operators. This builds on the classification presented in \cite{Marius}, see also the recent analysis performed in \cite{NewMarius}. 

\subsection{This paper}

Recently, the form factor programme has moved its first steps in $AdS_3$, with the intention of tackling the correlation functions. The paper \cite{AleSBurkhardtLePlat} has applied the hexagon approach (see references in \cite{AleSBurkhardtLePlat} for the advances made by this approach in $AdS_5$) to the $AdS_3$ case, and initiated the analysis of correlators in particular cases, especially those involving protected operators. In \cite{AleForm} we have entirely restricted to the massless sector and exploited the difference form to apply the more traditional method which Babujian et al devised for Sine-Gordon (see references within \cite{AleForm}), and conjectured integral expressions for the form factors up to three particles by solving the form-factor axioms. A general formula was contemplated for a general number of particles, but testing it is still beyond reach. 

In this paper we mostly remain in the context of massless modes, and explore an alternative idea due to \cite{Maillet,Maillet2}. This is based on the properties of Hopf algebras and in particular on the idea of factorising twists, which we will review in the text. The crude point is to try and make a change of basis whereby the calculation of form factors and correlation functions is remarkably simplified, and the complication is cast away within the change of basis. Such transformation makes use of Drinfeld twists and on the property of $R$-matrices to admit such factorising twists which reduce it to the identity. This is a feature which is deeply engrained in the quantum group origin of the $R$-matrix and its triangular decomposition, as we will discuss in what follows. Here we will move the very first few steps in this direction, as it turns out that our situation is, once again, harder than the ones traditionally investigated and solved via this technique. Nevertheless, we will be able to unveil useful properties and curious relations which will allow us to setup the programme in a very similar way to what is traditionally done. We will in fact be able to do this for massless modes in both $AdS_3$ and $AdS_2$. Along the way, we will also discover how to derive the massless quantum $AdS_2$ $R$-matrix from the Yangian universal $R$-matrix.  

The paper is strutured as follows: in section \ref{s2} we recall the notion of Hopf-algebra twists; in section \ref{s3} we recall the massless $AdS_3$ $S$-matrix; in section \ref{s4} we derive the factorising twist from the universal $R$-matrix of the $\alg{gl}(1|1)$ Yangian, discuss how the coproducts transform after we (un)do the twist, and tackle the three-site problem for the monodromy matrix; in section \ref{s5} we show how to recover a simple scalar product \cite{JuanMiguelAle} using the twist; in section \ref{specu} we speculate on the general case; in section \ref{s7} we make a connection with the free-fermion realisation \cite{Dublin}; in section \ref{s8} we make an extension to mixed-flux backgrounds; in section \ref{s9} we treat the $AdS_2$ case and show how to use the Yangian universal $R$-matrix to derive the massless $R$-matrix, alongside with a rather unusual version of the factorising twist. We then make some Conclusions and Acknowledgments.

\section{Hopf algebra twists\label{s2}}

We briefly recall here the notion of Hopf algebra twist \cite{Chari:1994pz}. Given a quasi-triangular Hopf algebra $\cal{A}$, where we denote by $\mathfrak{1}$ the unit with respect to the multiplication, with $\Delta$ the coproduct, with $\epsilon$ the counit and with $\cal{R}$ the universal $R$-matrix, we can obtain another quasi-triangular Hopf algebra by {\it twisting} it. This means that, if we can find an invertible element ${\cal{F}} \in {\cal{A}\otimes \cal{A}}$ satisfying
\begin{eqnarray}
(\epsilon \otimes \mathfrak{1}) \, {\cal{F}} = (\mathfrak{1} \otimes \epsilon) \, {\cal{F}} = \mathfrak{1}, \qquad ({\cal{F}} \otimes \mathfrak{1})(\Delta \otimes \mathfrak{1} ) {\cal{F}} = (\mathfrak{1} \otimes {\cal{F}}) (\mathfrak{1} \otimes \Delta) {\cal{F}}
\end{eqnarray}   
(with $\mathfrak{1}$ acting by multiplication in the above formulas), then the new coproduct and, respectively, new $R$-matrix
\begin{eqnarray}
\tilde{\Delta}(x) = {\cal{F}} \Delta(x) {\cal{F}}^{-1} \qquad \forall \, \, x \in {\cal{A}}, \qquad \tilde{{\cal{R}}} = {\cal{F}}^{op} {\cal{R}} {\cal{F}}^{-1},
\end{eqnarray}
define a new quasi-triangular Hopf algebra. 

The idea of a factorising twist correponds to finding a particular twists such that the new $R$-matrix is equal to $\mathfrak{1} \otimes \mathfrak{1}$. From the definition, this automatically implies that the old $R$-matrix can be factorised as
\begin{eqnarray}
{\cal{R}} =( {\cal{F}}^{op})^{-1} \, \mathfrak{1}\otimes \mathfrak{1} \, {\cal{F}} = ( {\cal{F}}^{op})^{-1}  {\cal{F}}.
\end{eqnarray}
Because the new $R$-matrix is equal to identity, the new coproduct must be co-commutative - although it can be highly non-trivial. In fact, one has
\begin{eqnarray}
\tilde{\Delta}^{op}(x) \tilde{{\cal{R}}} = \tilde{{\cal{R}}} \tilde{\Delta}(x) \qquad \forall \, \, x \in {\cal{A}}, \qquad \tilde{{\cal{R}}} = \mathfrak{1} \otimes \mathfrak{1} \qquad \leftrightarrow \qquad \tilde{\Delta}^{op} = \tilde{\Delta}.
\end{eqnarray}
For purely bosonic Hopf algebras a theorem by Drinfeld always guarantees the existence of a factorising twist under suitable assumptions -  see \cite{Maillet,Maillet2}, where a very detailed presentation of Hopf and quasi-Hopf algebras and their twists is also offered.

\section{The massless $AdS_3$ $S$-matrix\label{s3}}

In this paper we will only focus on the same-chirality massless $S$-matrix of the $AdS_3 \times S^3 \times T^4$ superstring, as it will be sufficient to illustrate our main results. We will also not be concerned at present with the scalar (so-called {\it dressing}) factor, but will instead focus merely on the suitably normalised matrix part. We will also make systematic use of the associated $R$-matrix. If the $S$-matrix is defined as
\begin{equation}
S : V_1 \otimes V_2 \longrightarrow V_2 \otimes V_1, \qquad S |v_\alpha (\gamma_1) \rangle \otimes |v_\beta (\gamma_2)\rangle = S^{\rho \sigma}_{\alpha \beta} (\gamma_1 - \gamma_2) |v_\rho (\gamma_2) \rangle \otimes |v_\sigma (\gamma_1)\rangle,\label{summat}
\end{equation}
the $R$-matrix is instead given by
\begin{eqnarray}
S = \Pi \circ R, \qquad R = \rho_1 \otimes \rho_2 \, ({\cal{R}}),
\end{eqnarray}
where $\Pi$ is the graded permutation on the states:
\begin{eqnarray}
\Pi |x\rangle \otimes |y\rangle = (-)^{[x][y]} \, |y\rangle \otimes |x\rangle,
\end{eqnarray}
where $[x]$ provides the fermionic number of the excitation $x$ and equals $0$ for bosons and $1$ for fermions. The notation $|v_\alpha(\gamma_i)\rangle$ denotes a basis of vectors of so-called {\it one-particle} representation: namely, we have a basis of the vector space $V_i$, which is of dimension $d_i$ (typically both $d_i=2$ for $i=1,2$ in this paper), hence $\alpha = 1,...,d_i$; such vector space carries an $\cal{A}$-representation $\rho_i:{\cal{A}} \to End(V_i)$, and such a representation is characterised by a particular parameter $\gamma_i$ (which is real for unitary representations). Such a parameter will play the role of the physical rapidity of the scattering states, but in general is to be understood as the eigenvalue of a particular Casimir element of $\cal{A}$ in the representation $\rho_i$. In (\ref{summat}) summation of repeated indices is understood. 

In the paper\footnote{We thank the anonymous referee for the pointing out a notational discrepancy in a previous version of the paper.} we will typically understand an italic $R$ symbol as denoting the representation $\rho_1 \otimes \rho_2$ of the corresponding $\cal{R}$ universal object (for instance, we will have $R_+$ by which we will mean $\rho_1 \otimes \rho_2 \, ({\cal{R}}_+)$, etc.). Likewise for the symbol $F$ (with respect to its corresponding $\cal{F}$).

We will also exploit the difference-form of the $S$-matrix in the appropriate pseudo-relativistic variables
\begin{eqnarray}
\gamma_i = \log \tan \frac{p_i}{4}, \qquad i=1,2,
\end{eqnarray}
$p_i$ being the momentum of the $i$-th particle.
The $R$-matrix then maps
\begin{equation}
R : V_1 \otimes V_2 \longrightarrow V_1 \otimes V_2.
\end{equation}
In the specific case we will study the matrix representation has been worked out in \cite{DiegoBogdanAle}, and reads  
\begin{equation}\label{eq:RLLlimtheta}
  \begin{aligned}
    &R |b\rangle \otimes |b\rangle\ = {} \, |b\rangle \otimes |b\rangle, \\
    &R |b\rangle \otimes |f\rangle\ = -\tanh\frac{\gamma}{2} |b\rangle \otimes |f\rangle + {\rm sech}\frac{\gamma}{2}|f\rangle \otimes |b\rangle, \\
    &R |f\rangle \otimes |b\rangle\ = {} \, {\rm sech}\frac{\gamma}{2} |b\rangle \otimes |f\rangle + \tanh\frac{\gamma}{2} |f\rangle \otimes |b\rangle, \\
    &R |f\rangle \otimes |f\rangle\ = - {} \, |f\rangle \otimes |f\rangle\,,
\end{aligned}
\end{equation}
where 
\begin{equation}
\label{eq:diff-form}
\gamma = \gamma_1 - \gamma_2\,
\end{equation}
and $(b,f)$ are a (boson, fermion) doublet, which we will conventionally number as $|b\rangle = |1\rangle$ and $|f\rangle = |2\rangle$ (not to be confused with the spaces $1$ and $2$ of the tensor product). We have also suppressed the $\gamma$ in the basis vectors to lighten up the notation (\ref{summat}) - here $d_i=2$ (in fact $1+1$) $\forall \, \, i=1,2$.

One can verify that this $R$-matrix satisfies
\begin{eqnarray}
R_{21} (-\gamma) R_{12}(\gamma) = 1_2 \otimes 1_2, \qquad R_{12}(\gamma) = R, \qquad R_{21}(\gamma) = perm \circ R,\label{bru}
\end{eqnarray}
where $1_2$ is the two-dimensional identity matrix. Property (\ref{bru}) is the manifestation of the {\it braiding unitarity} property of the $R$-matrix in the chosen two-dimensional representation. The map $perm$ is the graded permutation on the operators:
\begin{eqnarray}
perm (A \otimes B) = (-)^{[A][B]} B \otimes A, \qquad A,B : V_i \longrightarrow V_i, \qquad i=1 \, \, \mbox{or} \, \, 2,
\end{eqnarray}
and it is only defined for operators which are homogeneous with respect to the fermionic grading. In this paper we shall only consider operators which are homogeneous in this sense.
 
Practically, one can write 
\begin{eqnarray}
R = R_{12}(\gamma) = R^{abcd}(\gamma) E_{ab}\otimes E_{cd}, \qquad R_{21}(\gamma) = (-)^{([a]+[b])([c]+[d])} R^{abcd}(\gamma) E_{cd}\otimes E_{ab}.
\end{eqnarray}
In these formulas, all repeated indices are summed over the range $1,2$. $E_{ij}$ are the matrices with all zeroes except a $1$ in row $i$ and column $j$.
The $R$-matrix also has the curious property
\begin{eqnarray}
R_{12}^2(\gamma) = 1_2 \otimes 1_2. 
\end{eqnarray}

\section{$R$-matrix twist\label{s4}}

\subsection{Two sites from the Universal $R$-matrix}

The $R$-matrix twist is obtained by making use of the universal $R$-matrix formulation of the $\mathfrak{gl}(1|1)$ Yangian, which is the quantum supergroup underlying all $AdS_3$ $S$-matrices. We first review the Khoroshkin-Tolstoy construction which provides the universal $R$-matrix and which, when evaluated in specific representations, gives us the physical $R$-matrix described in the previous section. 

The first thing to do is to define the (super)Yangian of $\alg{gl}(1|1)$ in the so-called second realisation of Drinfeld's \cite{Drinfeld:1987sy}, see in particular \cite{Khoroshkin:1994uk,Koro2}\cite{unive0,unive}. In fact what admits a universal $R$-matrix is the  
(super)Yangian double $DY(\alg{gl}(1|1))$, which is generated by Chevalley-Serre elements ${}{e}_n$, ${}{f}_n$, ${}{h}_n$, ${}{k}_n$, $n\in\mathbbmss{Z}$, satisfying 
\begin{equation}
  \label{eq:Lie}
  \begin{gathered}
    \comm{{}{h}_0}{{}{e}_n} = -2{}{e}_n , \qquad
    \comm{{}{h}_0}{{}{f}_n} = +2{}{f}_n, \qquad
    \acomm{{}{e}_m}{{}{f}_n} = -{}{k}_{m+n} , \\
    \comm{{}{h}_m}{{}{h}_n} = 
    \comm{{}{h}_m }{{}{k}_n} =
    \comm{{}{k}_m }{{}{k}_n} = 
    \comm{{}{k}_m }{{}{e}_n} =
    \comm{{}{k}_m }{{}{f}_n} =
    \acomm{{}{e}_m}{{}{e}_n} = 
    \acomm{{}{f}_m}{{}{f}_n} = 0 , \\
    \comm{{}{h}_{m+1}}{{}{e}_n} - \comm{{}{h}_m}{{}{e}_{n+1}} + \acomm{{}{h}_m}{{}{e}_n} = 0, \qquad
    \comm{{}{h}_{m+1}}{{}{f}_n} - \comm{{}{h}_m}{{}{f}_{n+1}} - \acomm{{}{h}_m}{{}{f}_n} = 0.
  \end{gathered}
\end{equation}
Closely following \cite{Koro2}, one then needs to introduce the generating functions ({\it currents})
\begin{equation}
        K^{\pm}(t) = \mathfrak{1}\pm \sum_{\substack{n \ge 0 \\ n<0}} {}{k}_n t^{-n-1}, \qquad 
    H^{\pm}(t) = \mathfrak{1}\pm \sum_{\substack{n \ge 0 \\ n<0}} {}{h}_n t^{-n-1} .
\end{equation}
The universal $R$-matrix then reads
\begin{equation}
  \label{eq:rmatrix}
  {\cal{R}}={\cal{R}}_+{\cal{R}}_1{\cal{R}}_2{\cal{R}}_-, 
\end{equation}
where
\begin{equation}
  \begin{gathered}\label{pro}
    {\cal{R}}_+=\prod_{n\ge 0}\exp(- {}{e}_n\otimes {}{f}_{-n-1}),\qquad  
    {\cal{R}}_-=\prod_{n\ge 0}\exp({}{f}_n\otimes {}{e}_{-n-1}), \\ 
    \begin{aligned}
      {\cal{R}}_1 &= \prod_{n\ge 0} \exp \left\{ \mbox{res}_{t=z}\left[(-1) \frac{d}{dt}(\log H^+(t)) \otimes \log K^-(z+2n+1)\right]\right\}, \\
      {\cal{R}}_{2} &= \prod_{n\ge 0} \exp \left\{ \mbox{res}_{t=z}\left[(-1) \frac{d}{dt}(\log K^+(t))\otimes \log H^-(z+2n+1)\right]\right\}.
    \end{aligned}
  \end{gathered}
\end{equation}
The residue appearing in the above formula is given by
\begin{equation}
  \mbox{res}_{t=z}\left[A(t)\otimes B(z)\right] = \sum_k a_k\otimes b_{-k-1},
\end{equation}
after having re-expressed the currents in the form $A(t)=\sum_k a_k t^{-k-1}$, $B(z)=\sum_k b_k z^{-k-1}$.

The following ({\it evaluation}) representation $\rho$ satisfies all the relations \eqref{eq:Lie}, $\lambda$ being the so-called {\it evaluation} or {\it spectral} parameter:
\begin{equation}
  \rho{}({e}_n) = \lambda^n \, \rho(e_0), \qquad  
  \rho{}({f}_n) = \lambda^n \, \rho(f_0), \qquad
  \rho{}({k}_n) = \lambda^n \, \rho(k_0), \qquad 
 \rho{}({h}_n) = \lambda^n \rho(h_0),\label{onep}
\end{equation}
having set
\begin{eqnarray}
\rho(e_0) = a \, E_{12}, \qquad\rho (f_0) = d \, E_{21}, \qquad \rho(k_0) = - a d \, 1_2, \qquad \rho(h_0) = E_{11}-E_{22},
\end{eqnarray}
to match with the representation our physical $R$-matrix is written in, where both $a$ and $d$ can take indices $a_i$ and $d_i$, $i=1,2$, to denote which space they pertain to. The parameters $a_i$ and $d_i$ are a priori complex numbers which determine the particular representation of interest, and will be set to specific values in order to reproduce particular physical situations occuring in string theory. The same holds for $\lambda$, which acquires the index $\lambda_i$, $i=1,2$. Such a representation can equivalently be reformulated by defining an evaluation map from $DY(\mathfrak{gl}(1|1))$ onto the universal enveloping algebra $U(\mathfrak{gl}(1|1))$ of the Lie superalgebra $\mathfrak{gl}(1|1)$, and then provide the appropriate representation of $U(\mathfrak{gl}(1|1))$.

Because of the fermionic nature of the root generators, the root part of the universal $R$-matrix naturally truncates to
\begin{equation}
  \begin{aligned}
    R_+ &= 1_2 \otimes 1_2-\sum_{n\geq0}{}\rho_1({e}_n)\otimes {}\rho_2({f}_{-n-1})=1_2 \otimes 1_2+\frac{\rho_1(e_0)\otimes \rho_2(f_0)}{\lambda_1 - \lambda_2},\\
    R_- &= 1_2 \otimes 1_2+\sum_{n\geq0}{}\rho_1({f}_n)\otimes {}\rho_2({e}_{-n-1})=1_2 \otimes 1_2-\frac{\rho_1(f_0)\otimes \rho_2(e_0)}{\lambda_1 - \lambda_2}.
\end{aligned}
\end{equation}
Specifically one can write in matrix form
\begin{eqnarray}
R_+ = \begin{pmatrix}1&0&0&0\\0&1&\frac{a_1 d_2}{\lambda}&0\\0&0&1&0\\0&0&0&1\end{pmatrix}, \qquad R_- = \begin{pmatrix}1&0&0&0\\0&1&0&0\\0&\frac{a_2 d_1}{\lambda}&1&0\\0&0&0&1\end{pmatrix},
\end{eqnarray}
with\footnote{\label{confu}It is mildly confusing to call $\lambda$ again the difference: we hope that the context will always clarify between $\lambda$ the difference and $\lambda$ in the one-particle representation (\ref{onep}), where it really should be $\lambda_i$.} $\lambda = \lambda_1 - \lambda_2$.
When it comes to the Cartan generators, in an appropriate domain of convergence one finds
\begin{equation}
  -\frac{d}{dt}\log H^+ = \sum_{m=1}^{\infty} \left\{{\lambda^m} -{(\lambda-{}\rho_1({h}_0))^m}\right\} t^{-m-1}
\end{equation}
and
\begin{multline}
  \log K^-(z+2n+1) = \log K^-(2n+1) + \\
  +\sum_{m=1}^{\infty} \left\{\frac{1}{(\lambda-1-2n)^m} -\frac{1}{(\lambda-1-2n-{}\rho_2({k}_0))^m}\right\} \frac{z^{m}}{m},
\end{multline}
and similarly for $R_2$. Since the generators $h_0$ and $k_0$ appearing in these expressions are diagonal, the way they appear in the functions is of no concern and one can obtain expressions for each diagonal entry. Taking the residue and resumming these expressions is still a lengthy procedure, which can be found described for instance in \cite{unive}. The final result reads
\begin{eqnarray}
\label{R1e2}
R_1 = \Omega_1 \begin{pmatrix}1&0&0&0\\0&\frac{\lambda - a_2 d_2}{\lambda}&0&0\\0&0&1&0\\0&0&0&\frac{\lambda - a_2 d_2}{\lambda}\end{pmatrix}, \qquad R_2 = \Omega_2\begin{pmatrix}1&0&0&0\\0&1&0&0\\0&0&\frac{\lambda}{\lambda + a_1 d_1}&0\\0&0&0&\frac{\lambda}{\lambda + a_1 d_1}\end{pmatrix}, 
\end{eqnarray}
where $\Omega_i$, $i=1,2$, are two scalar factors built out of gamma matrices which we are ignoring as irrelevant to our present discussion.

We now select the representation parameters as
\begin{eqnarray}
a_j = \sqrt{i(x^-_j - x^+_j)}, \qquad d_j = \frac{x^+_j - x^-_j}{i a_j}, \qquad \lambda_j = - i x^-_j + const, \qquad j=1,2,
\end{eqnarray}
to obtain the $AdS_3$ $R$-matrix in Zhukovsky variables. We refer to \cite{Foundations}, in particular section 3.23, for one description of a version of this parameterisation as it first originated in this context in a related problem (the $AdS_5 \times S^5$ superstring) and the connection with the Zhukovski map, which justifies its name. Reference \cite{unive} is then more closely related to our present case, especially section 6, and references therein. The constant in the parameterisation for $\lambda_i$ is irrelevant and will cancel in the diffference $\lambda$. After evaluating the $R$-matrix in this parameterisation, we finally take the massless limit in the same-chirality setting 
\begin{eqnarray}
x^\pm_i = e^{\pm i \frac{p_i}{2}}
\end{eqnarray}
and obtain exactly the matrix $R_{12}(\gamma)$, up to a simple twist\footnote{This simple twist is described in \cite{unive}, see also \cite{Foundations}, and corresponds to a choice of frame factors. Such an additional twist could easily be incorporated but it does not add anything to our discussion, therefore we have chosen to neglect it.}, provided we also change the sign of $\gamma$ in the final matrix\footnote{The universal $R$-matrix does not know a priori which ordering of the momenta we choose, so we believe that in principle it should not be a problem to change $\gamma \to -\gamma$ (equivalently $\gamma_1$ with $\gamma_2$) in the final formula, as long as we do it systematically all along. We can in fact take the point of view that the universal $R$-matrix, for our purposes, just gives us a systematic recipe to obtain the (twist) matrix in representations from an abstract framework  - if the recipe involves an additional simple systematic operation at the end, then the recipe will have served its present purpose.}. 

As explained in \cite{korotwi}, the above formula for the universal $R$-matrix is already designed to be factorised, and we can simply read off the factorising twist:
\begin{eqnarray}
\label{two}
{\cal{F}}_{12}(\gamma) = {\cal{R}}_2 \, {\cal{R}}_-, \qquad {\cal{F}}_{21}^{-1}(-\gamma) = {\cal{R}}_+ \, {\cal{R}}_1.
\end{eqnarray}
One can also brute-force verify using the actual $4 \times 4$ matrices that the two equations in (\ref{two}) are compatible with each other in the representation $\rho_1 \otimes \rho_2$. One specifically finds (again having changed $\gamma \to -\gamma$ in the final step)
\begin{eqnarray}
F_{12}(\gamma) = E_{11}\otimes E_{11} + E_{11}\otimes E_{22} + \tanh \frac{\gamma}{2} (E_{22}\otimes E_{11} + E_{22}\otimes E_{22}) + \mbox{sech}\frac{\gamma}{2} E_{21}\otimes E_{12},
\end{eqnarray}
or
\begin{eqnarray}
F_{12}(\gamma) = \begin{pmatrix}1&0&0&0\\0&1&0&0\\0&\mbox{sech}\frac{\gamma}{2}&\tanh \frac{\gamma}{2}&0\\0&0&0&\tanh \frac{\gamma}{2}\end{pmatrix}, \qquad F_{21}^{-1}(-\gamma) = \begin{pmatrix}1&0&0&0\\0&-\mbox{coth}\frac{\gamma}{2}&\mbox{csch}\frac{\gamma}{2}&0\\0&0&1&0\\0&0&0&-\mbox{coth}\frac{\gamma}{2}\end{pmatrix},
\end{eqnarray}
where we denote ${.}_{21} = {.}^{op}$ indifferently.
The fact that the twist is a lower triangular matrix literally corresponds to the fact that the Khoroshkin-Tolstoy formula highlights the block-triangular decomposition of the $R$-matrix. This is in turn associated with the Chevalley-Serre (Drinfeld's second) presentation of the Yangian relations based on the positive and negative root- and Cartan element decomposition. Notice that (\ref{two}) combined with (\ref{pro}) provides a universal expression for the twist.

We also remark that, in the specific representation we are using, because of the peculiarity of the massless $AdS_3$ $R$-matrix which satisfies $R_{21}(-\gamma)=R_{12}(\gamma)$ (and hence, as we pointed out earlier, $R^2(\gamma) = 1_2 \otimes 1_2$ because of braiding unitarity) \cite{JoakimAle}, we have that equally
\begin{eqnarray}
R = F_{21}^{-1}(-\gamma) F_{12}(\gamma)= F_{12}^{-1}(\gamma) F_{21}(-\gamma).
\end{eqnarray}

\subsection{Twisted Symmetries}

Let us define 
\begin{eqnarray}
\Delta_\rho \equiv [\rho_1 \otimes \rho_2]
 \circ \Delta,
\end{eqnarray}
and similarly for the other symbols $\delta,\tilde{\Delta} $ which we will use to denote different coproduct maps.
We will also use the notation 
\begin{eqnarray}
Q \equiv e_0, \qquad Q^\dagger \equiv f_0, \qquad b_n \equiv h_n,
\end{eqnarray}
in order to connect with the traditional notation utilised in the AdS/CFT literature. 

To inspect what the coproduct of the simplest (level-zero) supercharges becomes after the twist, we can begin by focusing on the ${\cal{N}}=2$ relativistic supersymmetry which the $R$-matrix inherits simply by virtue of being of the same form as its relativistic limit. One can in fact directly verify \cite{Dublin} that the (massless- relativistic- like) supercharges
\begin{eqnarray}
\Delta_\rho(Q) = e^{\frac{\gamma_1}{2}}E_{12} \otimes 1_2 + 1_2 \otimes e^{\frac{\gamma_2}{2}} E_{12}, \qquad \Delta_\rho(Q^\dagger) = e^{\frac{\gamma_1}{2}}E_{21} \otimes 1_2 + 1_2 \otimes e^{\frac{\gamma_2}{2}} E_{21}
\end{eqnarray} 
are (already co-commutative) symmetries of the $R$-matrix: $\Delta_\rho^{op}(Q)R=R\Delta_\rho(Q)$ and  $\Delta_\rho^{op}(Q^\dagger)R=R\Delta_\rho(Q^\dagger)$. The twist should maintain them co-commutative, since the new $R$-matrix equals the identity. By direct computation we find, with $\gamma = \gamma_1 - \gamma_2$,
\begin{eqnarray}
&&\tilde{\Delta}_\rho(Q) = F_{12}(\gamma) \Delta_\rho(Q)F_{12}^{-1}(\gamma) = \mbox{coth}\frac{\gamma}{2} \Big[-e^{\frac{\gamma_2}{2}} 1_2\otimes E_{12} + e^{\frac{\gamma_1}{2}} E_{12}\otimes 1_2\Big],\nonumber\\
&&\tilde{\Delta}_\rho(Q^\dagger) = F_{12}(\gamma) \Delta_\rho(Q^\dagger)F_{12}^{-1}(\gamma) = \Delta_\rho(Q^\dagger),\nonumber
\end{eqnarray}
both satisfying the co-commutativity condition: in the case of $\tilde{\Delta}_\rho(Q)$ the prefactor changes sign under ${}^{op}$, and so does the bracket; in the case of $\tilde{\Delta}(Q^\dagger)$, we have that it equals $\Delta_\rho(Q^\dagger)$ which was of course already co-commutative.

In \cite{Dublin} it was also shown (see discussion around formula (4.60) there) how the more traditional AdS/CFT braided coproducts, in the case of massless $AdS_3$, reduce to combinations\footnote{It was shown in \cite{JoakimAle} that for massless $AdS_3$ there are actually two inequivalent coproducts one can use, one more standard and the other one only existing thanks to the particular form of the massless dispersion relation. Both these versions are reported in formula (4.55) in \cite{Dublin}.} of the above ${\cal{N}}=2$ one and another symmetry, let us call it $\delta q$
\begin{eqnarray}
\label{that}
\delta_\rho q = e^{\frac{\gamma_1}{2}+\gamma_2} E_{12} \otimes 1_2 - e^{\frac{\gamma_2}{2}+\gamma_1} 1_2 \otimes E_{12} 
\end{eqnarray}
(and the hermitian conjugate obtained replacing $E_{12}$ by $E_{21}$). One can verify by brute force that (\ref{that}) is a (this time non co-commutative) symmetry. The action of the twist turns this into
\begin{eqnarray}
\widetilde{\delta_\rho q} = \mbox{coth}\frac{\gamma}{2} \Big[-e^{\gamma_1+\frac{\gamma_2}{2}} 1_2\otimes E_{12} + e^{\gamma_2+\frac{\gamma_1}{2}} E_{12}\otimes 1_2\Big],
\end{eqnarray}
which is indeed co-commutative (again by compensation of two minuses under the ${}^{op}$ operation, one from the pre-factor and one from the bracket). The hermitian conjugate is instead turned into
\begin{eqnarray}
\widetilde{\delta_\rho q^\dagger} = -e^{\gamma_1+\frac{\gamma_2}{2}} 1_2\otimes E_{21} - e^{\gamma_2+\frac{\gamma_1}{2}} E_{21}\otimes 1_2,
\end{eqnarray}
which is co-commutative by direct inspection.

Combining these results and the formula provided in \cite{Dublin}, we can write the (two inequivalent) massless $AdS_3$ coproducts after the twist as the following co-commutative combinations\footnote{We repeat here the following definition from \cite{Dublin} for ease of reading:
\begin{eqnarray}
Q_\pm \equiv \sqrt{\sin \frac{p_1}{2}}e^{\pm i \frac{p_2}{4}} E_{12} \otimes 1_2 + \sqrt{\sin \frac{p_2}{2}}e^{\mp i \frac{p_1}{4}} 1_2 \otimes E_{12},\nonumber
\end{eqnarray}
with $p_i \in (0,\pi)$.
Despite the use of the letter `$Q$', the combinations $Q_\pm$ above are operators acting on the space $V_1 \otimes V_2$. In physical parlance they were obtained simply as brute-force symmetries of the $S$-matrix - in an equivalent sense we mean that they satisfy 
\begin{eqnarray}
Q_\pm^{op} R = R Q_\pm \nonumber
\end{eqnarray} 
by brute force computation.
}:
\begin{eqnarray}
\tilde{Q}_{\pm} = g_{12} \, \mbox{coth}\frac{\gamma}{2} \Big[(1\pm i e^{\gamma_2}) e^{\frac{\gamma_1}{2}}E_{12} \otimes 1_2-(1\pm i e^{\gamma_1}) 1_2 \otimes e^{\frac{\gamma}{2}}E_{12}\Big]
\end{eqnarray}
where $g_{12}$ is a scalar function symmetric under the exchange of $1$ and $2$, hence transparent to the symmetry condition (and trivially co-commutative itself), and the coth again changing sign under ${}^{op}$ and compensating the bracket. In the case of the hermitian conjugate we get after the twist
the co-commutative combinations
\begin{eqnarray}
\tilde{Q}_{\pm}^\dagger = g_{12} \, \Big[(1\mp i e^{\gamma_2}) e^{\frac{\gamma_1}{2}}E_{21} \otimes 1_2+(1\mp i e^{\gamma_1}) 1_2 \otimes e^{\frac{\gamma}{2}}E_{21}\Big].
\end{eqnarray}

AdS/CFT integrability also admits a so-called secret symmetry \cite{unive0,secret,segreto}. In the $AdS_3$ context there is a level zero version of this symmetry (at odds with $AdS_5$ and $AdS_4$), namely
\begin{eqnarray}
\Delta_\rho(b_0) = (E_{11}-E_{22}) \otimes 1_2 + 1_2 \otimes (E_{11}-E_{22}),
\end{eqnarray}
with $F$ the fermion number. There is then a level one Yangian correspondent, which in the massless case reduces to a ``conformal'' Yangian symmetry \cite{Ale}:
\begin{eqnarray}
\Delta_\rho(b_1) = e^{\frac{\gamma_1 + \gamma_2}{2}} \Big[E_{12} \otimes E_{21} + E_{21} \otimes E_{12}\Big].
\end{eqnarray}
After the twist, they become
\begin{eqnarray}
\tilde{\Delta}_\rho(b_0) = 2 E_{11} \otimes E_{11} - 2 E_{22} \otimes E_{22},
\end{eqnarray}
which is clearly co-commutative, and, respectively ($\gamma$ always being $\gamma_1 - \gamma_2$),
\begin{eqnarray}
\tilde{\Delta}_\rho(b_1)= \mbox{coth}\frac{\gamma}{2}\Big[\frac{2 e^{\gamma_1 + \gamma_2}}{e^{\gamma_1} + e^{\gamma_2}} (E_{11} \otimes E_{22} - E_{22} \otimes E_{11}) + e^{\frac{\gamma_1 +\gamma_2}{2}} (E_{12} \otimes E_{21} + E_{21} \otimes E_{12})\Big].
\end{eqnarray}
The latter is also co-commutative, considering that the prefactor changes sign, and also the bracket (recalling that both $E_{12}$ and $E_{12}$ are fermionic generators), under the ${}^{op}$ operation.

\subsection{Three sites}

The generalisation to three sites (and then an arbitrary number) is made difficult by the fact that the coproduct, although known in principle for the Yangian of $\alg{gl}(1|1)$, is not easy to handle. In particular, it is given in closed form on the currents, while on the modes it is obtained in principle by matching Laurent expansions, making the explicit expression bulky to manage. Nevertheless, we have found a concrete matrix which provides the answer for three sites.

For three sites, we require to solve the following factorisation problem:
\begin{eqnarray}
F_{321}(\gamma_3,\gamma_2,\gamma_1) R_{123}(\gamma_1,\gamma_2,\gamma_3) = F_{123}(\gamma_1,\gamma_2,\gamma_3),
\end{eqnarray} 
where
\begin{eqnarray}
R_{123}(\gamma_1,\gamma_2,\gamma_3) = R_{12}(\gamma_1-\gamma_2)R_{13}(\gamma_1-\gamma_3)R_{23}(\gamma_2-\gamma_3),
\end{eqnarray}
with the familiar assignment of indices for the triple tensor product. The factorising twist $F_{123}$ ought to be an invertible, lower-triangular matrix. It is not difficult to find an explicit solution to this condition, which is parameterised by four unknown functions $f_k(\gamma_1,\gamma_2,\gamma_3)$, $k=1,2,3,4$, all of which must be \underline{symmetric} functions under the exchange of $\gamma_1$ and $\gamma_3$: 
\begin{eqnarray}
F_{123}(\gamma_1,\gamma_2,\gamma_3) = F_{12}(\gamma_1-\gamma_2) \Big[D_1 + D_2 \cdot E_{21}\otimes 1_2 \otimes E_{12} + D_3 \cdot 1_2 \otimes E_{21} \otimes E_{12}\Big],\label{ansaz}
\end{eqnarray}
the $D_q$, $q=1,2,3$, being diagonal matrices
\begin{eqnarray}
\label{24}
D_1 = \mbox{diag}(\tilde{d}_1,...,\tilde{d}_8), \qquad D_2 = \mbox{diag}(\tilde{e}_1,...,\tilde{e}_8), \qquad
D_3 = \mbox{diag}(\tilde{f}_1,...,\tilde{f}_8),
\end{eqnarray} 
satisfying
\begin{eqnarray}
&&\tilde{d}_7 = f_4 \, \mbox{coth}\frac{\gamma_1-\gamma_2}{2} \, \tanh \frac{\gamma_2-\gamma_3}{2},\nonumber\\
&&\tilde{f}_7 = f_4 \, \mbox{sech} \frac{\gamma_2-\gamma_3}{2},\nonumber\\
&&\tilde{e}_7 = -f_4 \, \mbox{coth}\frac{\gamma_1-\gamma_2}{2} \, \mbox{csch} \frac{\gamma_1-\gamma_3}{2}, \qquad \tilde{d}_6 = f_4 \, \mbox{coth}\frac{\gamma_1-\gamma_3}{2},\nonumber\\
&&\tilde{d}_5 = \tilde{d}_3 = \tilde{f}_3  \,  \tanh \frac{\gamma_2-\gamma_3}{2}, \qquad \tilde{d}_1 = f_2, \qquad \tilde{d}_4 = f_4 \, \mbox{coth}\frac{\gamma_1-\gamma_3}{2}, \nonumber\\
&&\tilde{e}_5 = f_3  \, \mbox{coth}\frac{\gamma_1-\gamma_2}{2} \, \mbox{csch}\frac{\gamma_1-\gamma_3}{2} \,   \tanh \frac{\gamma_2-\gamma_3}{2}, \nonumber\\
&&\tilde{d}_2 = f_3 \tanh \frac{\gamma_1-\gamma_2}{2} \, \mbox{coth}\frac{\gamma_1-\gamma_3}{2}, \nonumber\\
&&\tilde{d}_8 = f_1 \, \mbox{coth} \frac{\gamma_1-\gamma_2}{2} \, \tanh \frac{\gamma_1-\gamma_3}{2}, \qquad \tilde{f}_3 = f_3 \, \mbox{sech} \frac{\gamma_2-\gamma_3}{2}.\label{soluzw}
\end{eqnarray}
By convention, if no argument is given to a function of three variables in these formulas (for instance, say, just writing $\tilde{d}_4$), then it is understood to be in the 123 ordering, namely $(\gamma_1,\gamma_2,\gamma_3)$. Notice that not all the $24$ parameters (\ref{24}) appear in (\ref{soluzw}). This is because in reality they do not feature in the ansatz itself, since multiplying by the $E_{12}$s and $E_{21}$s effectively eliminates a lot of parameters (making them redundant).   

Notice that the solution we have found in this way is automaticaly lower triangular given the ansatz we started from, and it is invertible as long as all the functions $f_k$ are non-zero. A justification for the ansatz (\ref{ansaz}) will be provided in section \ref{specu}.

At this point we can verify one of the theorems of \cite{Maillet} on three sites. Since we have that
\begin{eqnarray}
R_{123}(\gamma_1,\gamma_2,\gamma_3) T_1(\gamma_1)T_2(\gamma_2)T_3(\gamma_3) = T_3(\gamma_3)T_2(\gamma_2)T_1(\gamma_1)R_{123}(\gamma_1,\gamma_2,\gamma_3),
\end{eqnarray}
where $T_i$ is the monodromy matrix with auxiliary space $i$ (whose entry act on the quantum space of a chain of $L$ sites), then the factorisation property implies (suppressing the arguments for simplicity, as they always follow the assignment of auxiliary spaces)
\begin{eqnarray}
F_{321}^{-1}F_{123}T_1 T_2 T_3 = T_3 T_2 T_1 F_{321}^{-1}F_{123}.\label{virtue}
\end{eqnarray}
Therefore, by defining
\begin{eqnarray}
\tilde{T}_{123} = F_{123} T_1 T_2 T_3 F_{123}^{-1},
\end{eqnarray}
we see that
\begin{eqnarray}
\tilde{T}_{123} = \tilde{T}_{321}
\end{eqnarray}
by virtue of (\ref{virtue}).

\section{A simple scalar product\label{s5}}

In this section we show how a simple scalar product can be reproduced by means of the factorising twist. This is in a perfect alignment with the philosophy of \cite{Maillet}.

Let us consider the quantity
\begin{eqnarray}
\label{scala}
\langle 0| C(\gamma_1) B(\gamma_2)|0\rangle,
\end{eqnarray}
which is related to the off-shell scalar product of one-magnon states\footnote{The precise relationship is carefully detailed in \cite{JuanMiguelAle}, section 4.1.}. The operators $B$ and $C$ are the creation and annihilation operators of the algebraic Bethe ansatz: namely, the monodromy matrix with auxiliary site $0$ can be written as
\begin{eqnarray}
T_0 = E_{11} \otimes A + E_{12} \otimes B + E_{21} \otimes C + E_{22} \otimes D,
\end{eqnarray} 
with the $E_{ij}$ acting on the space $0$ (depending on an auxiliary parameter $\gamma_0$) and $A,B,C,D$ acting on the quantum space of a chain of $L$ sites. The pseudovacuum $|0\rangle$ is taken as the state with all bosons along the chain.

The scalar product (\ref{scala}) is simple enough to be computable from the RTT relations in one move. Since the pseudovacuum satisfies
\begin{eqnarray}
C(\gamma_0)|0\rangle = 0, \qquad A(\gamma_0)|0\rangle = \alpha(\gamma_0) |0\rangle, \qquad D(\gamma_0)|0\rangle = \delta(\gamma_0) |0\rangle 
\end{eqnarray}
(having suppressed the dependence on the quantum spaces with inhomogeneities $\gamma_m$, $m=1,...,L$), we can simply use the particular RTT relation
\begin{eqnarray}
[C(\gamma_1),B(\gamma_2)] = \mbox{csch}(\gamma_1 - \gamma_2)\Big(D(\gamma_2)A(\gamma_1) - D(\gamma_1)A(\gamma_2)\Big),
\end{eqnarray}
to obtain
\begin{eqnarray}
\langle 0| C(\gamma_1) B(\gamma_2)|0\rangle = \mbox{csch}(\gamma_1 - \gamma_2)\Big(\alpha(\gamma_1)\delta(\gamma_2) - \alpha(\gamma_2)\delta(\gamma_1)\Big).\label{coin}
\end{eqnarray}
This can be made completely explicit by recalling that \cite{JuanMiguelAle}
\begin{eqnarray}
\alpha(\gamma)=1, \qquad \delta(\gamma) = \prod_{m=1}^L \tanh \frac{\gamma - \gamma_m}{2}.
\end{eqnarray}

The same result can be obtained by means of the factorising twist. Indeed one can extract the operators $C$ and $B$ by (super)tracing the monodromy matrix in its auxiliary space:
\begin{eqnarray}
C = \mbox{str}_0 \, E_{12} \otimes 1_N \cdot T_0, \qquad B = - \mbox{str}_0 \, E_{21} \otimes 1_N \cdot T_0,
\end{eqnarray}
where str denotes the supertrace and $1_{2^N} \equiv 1_2 \otimes 1_2 \otimes ... \otimes 1_2$ ($N$ times, as many as the number of physical, namely non-auxiliary, spaces of the monodromy matrix). One can then rely on another theorem in \cite{Maillet}, whose exact assumptions we satisfy in our case. From the very definition of the factorising twist we have in fact (suppressing arguments)
\begin{eqnarray}
R_{12} T_1 T_2 = T_2 T_1 R_{12} \qquad \leftrightarrow \qquad F_{21}^{-1} F_{12} T_1 T_2 = T_2 T_1 F_{21}^{-1} F_{12},
\end{eqnarray}
hence
\begin{eqnarray}
F_{12} T_1 T_2 F_{12}^{-1} = F_{21} T_2 T_1 F_{21}^{-1},\label{by}
\end{eqnarray}
which means that $\tilde{T}_{12} = F_{12} T_1 T_2 F_{12}^{-1}$ is symmetric\footnote{In a sense the relations
\begin{eqnarray}
\tilde{T}_{12} = \tilde{T}_{21}\nonumber
\end{eqnarray}
are the abstract RTT relations after the twist. A particular representation (of course not the only one) is  
\begin{eqnarray}
\tilde{T}_{12} = \tilde{T}_1 \tilde{T}_2, \qquad \tilde{T}_i = \prod_{j=1}^N \tilde{R}_{ij} = 1_2 \otimes 1_2 \otimes ... \otimes 1_2 \, \, \, \, \, (N+1 \, \, \mbox{times}), \qquad i=1,2.\nonumber
\end{eqnarray}  
} under the exchange of $1$ and $2$. Moreover, since $F_{12}$, $F_{12}^{-1}$ and $\langle 0| T_1 T_2|0\rangle$ are all lower trianguar\footnote{One can see that 
\begin{eqnarray}
\langle 0| T_1 T_2|0\rangle = (-)^{([i]+[j])([k]+[l])} E_{ij} \otimes E_{kl} \langle 0| T_1^{ij} T_2^{kl}|0\rangle \nonumber
\end{eqnarray}
while
\begin{eqnarray}
\langle 0| T_2 T_1|0\rangle = E_{ij} \otimes E_{kl} \langle 0| T_1^{kl} T_2^{ij}|0\rangle, \nonumber
\end{eqnarray}
so if one is a lower triangular matrix in the tensor product of the two auxiliary space $1$ and $2$, the other is upper triangular. One concludes by recalling that $T^{11} = A, \, T^{12}=B, \, T^{21}=C, \, T^{22}=D$.
}, we also have that 
\begin{eqnarray}
F_{12} \langle 0| T_1 T_2|0\rangle F_{12}^{-1}\label{1}
\end{eqnarray}
is lower triangular. Likewise, 
\begin{eqnarray}
F_{21} \langle 0| T_2 T_1|0\rangle F_{21}^{-1}\label{2}
\end{eqnarray}
is upper triangular. But since (\ref{1}) and (\ref{2}) are equal - by sandwiching (\ref{by}) between two pseudo-vacua - then they must be both equal to a diagonal matrix in the tensor product of the auxiliary spaces $1$ and $2$. 
In addition, these similarities must both equal the diagonal part of $\langle 0| T_2 T_1|0\rangle$. Based on this one therefore concludes that  
\begin{eqnarray}
\label{uaggio}
\langle 0|T_1 T_2|0\rangle = F_{12}^{-1} \Big[\langle 0|T_1|0\rangle \otimes \langle 0|T_2|0\rangle\Big] F_{12}, \qquad \langle 0|T_i|0\rangle = \begin{pmatrix}\alpha(\gamma_i)&0\\0&\delta(\gamma_i)\end{pmatrix}, \qquad i=1,2.
\end{eqnarray}
 
At this point one can easily see that
\begin{eqnarray}
\langle 0| C(\gamma_1) B(\gamma_2)|0\rangle = - \mbox{str}_{12} E_{12} \otimes E_{21} \langle 0|T_1 T_2|0\rangle,
\end{eqnarray}
since the quantum and the auxiliary spaces do not interfere, hence using (\ref{uaggio}) and (graded) cyclicity of the supertrace one obtains
\begin{eqnarray}
\langle 0| C(\gamma_1) B(\gamma_2)|0\rangle = - \mbox{str}_{12} \Big(F_{12} \cdot E_{12} \otimes E_{21} \cdot F_{12}^{-1} \cdot [\alpha(\gamma_1) E_{11} + \delta(\gamma_1) E_{22}]\otimes [\alpha(\gamma_2) E_{11} + \delta(\gamma_2) E_{22}]\Big), \nonumber \\\label{rhs}.
\end{eqnarray}
We have denoted $\mbox{str}_{12} = \mbox{str}_{1} \otimes \mbox{str}_{2}$, which is the natural supertrace on the tensor product space.
We have at this point all the ingredients from the previous sections, so it is a simple exercise multiplying out these matrices to prove that the r.h.s. of (\ref{rhs}) exactly coincides with the r.h.s. of (\ref{coin}). 

This corresponds to the statement \cite{Maillet} that the factorising twists provide a change of basis which diagonalise the scalar products of Bethe states. 

\section{Speculation on a general formula and form factors\label{specu}}

The natural question is to try and extend the result from three to an arbitrary number of sites. In the case of the models dealt with in \cite{Maillet}\cite{Maillet2}, a special relationship between the entries of the twist matrix and the $R$-matrix allowed a particular ansatz to be made for the general case. Unfortunately such relationship is slightly different here, enough to prevent that ansatz to go through. Specifically, one can prove by direct calculation that
\begin{eqnarray}
F_{12}(\gamma) = 1_2 \otimes E_{22} + 1_2 \otimes E_{11} \cdot R_{12}(\gamma) + \tau,\label{nice}
\end{eqnarray}
where $\tau$ is the (undesired) term $\tau = (\tanh \frac{\gamma}{2}-1) E_{22} \otimes E_{22}$. This extra term is undesired precisely because it does not behave well (or at least in a predictable way) under $\Delta \otimes \mathfrak{1} $ - while all other terms in (\ref{nice}) do. In fact, it is known how $\Delta$ acts on the unit element of $\cal{A}$, and how $\Delta \otimes \mathfrak{1} $ acts on $\cal{R}$ - one simply uses the quasi-triangular condition $(\Delta \otimes \mathfrak{1} ) {\cal{R}} = {\cal{R}}_{13}{\cal{R}}_{23}$. In the end this means that we cannot reduce the general factorising twist to a product of $R$-matrices. 

As mentioned earlier, using the coproduct directly on the universal expression for the twist is not easily managed either. One could in fact in principle apply the formula ${\cal{F}}_{123} = {\cal{F}}_{12}(\Delta \otimes \mathfrak{1} ) {\cal{F}}$, but in practice this is not easy to calculate. We can at least provide here a justification for the three-site ansatz (\ref{ansaz}). By using the homomorphism property of the coproduct we have
\begin{eqnarray}
(\Delta \otimes \mathfrak{1} ) {\cal{F}} = (\Delta \otimes \mathfrak{1} ) {\cal{R}}_2 \cdot (\Delta \otimes \mathfrak{1} ) {\cal{R}}_-,
\end{eqnarray} 
and using the universal expressions (\ref{pro}) we obtain
\begin{eqnarray}
&&(\Delta \otimes \mathfrak{1} ) {\cal{R}}_- = \prod_{n\geq 0} \exp \Delta(f_n) \otimes e_{-n-1}\nonumber\\
&&(\Delta \otimes \mathfrak{1} ){\cal{R}}_{2} = \prod_{n\ge 0} \exp  \mbox{res}_{t=z}\left[(-1) \frac{d}{dt}(\log \Delta(K^+(t))\otimes \log H^-(z+2n+1)\right].
\end{eqnarray}
Given the expected schematic structure \cite{Koro2} 
\begin{eqnarray}
\Delta(K^+(u)) =  \mbox{Cartan current} \otimes \mbox{Cartan current}, \qquad
\Delta(f) = \mathfrak{1} \otimes f + f \otimes \mbox{Cartan current},\nonumber
\end{eqnarray} 
(where $f$ schematically denotes generators of the type $f_n$) we can therefore see that in the representation of interest the ansatz (\ref{ansaz}) corresponds to the possible terms deriving from the universal formula.

\section{Free fermion realisation\label{s7}}

In \cite{Dublin} it was shown how to recast the $AdS_3$ massless same-chirality $S$-matrix in terms of free-fermion oscillators, in a way that vastly simplifies its appearance. One first introduces fermionic operators
\begin{eqnarray}
c_1 = E_{12}\otimes 1_2, \qquad c_2 = 1_2 \otimes E_{12}, \qquad c_1^\dagger = E_{21}\otimes 1_2, \qquad c_2^\dagger = 1_2 \otimes E_{21},
\end{eqnarray}
obeying canonical anti-commutation relations $\{c_i,c^\dagger_j\}= \delta_{ij}$ (and zero otherwise\footnote{In the mapping to the tensor-product matrices, the oscillator-language `$1$' corresponds to $1_2 \otimes 1_2$.}), such that the oscillator vacuum is $|1\rangle\otimes |1\rangle$. A canonical transformation is then applied:
\begin{eqnarray}
c_1 = \cos \alpha \, \eta_1 - \sin \alpha  \, \eta_2, \qquad c_2 = \sin \alpha \, \eta_1 + \cos \alpha \, \eta_2 
\end{eqnarray}
(together with the hermitian conjugates). This transformation is designed in such a way that the new generators $\eta_i, \eta^\dagger_i$, $i=1,2$, still satisfy canonical anti-commutation relations $\{\eta_i,\eta^\dagger_j\}= \delta_{ij}$ (and zero otherwise). The choice of branch for the transformation parameter is determined by setting 
\begin{eqnarray}
\mbox{cot}\alpha = - e^{-\frac{\theta_1 - \theta_2}{2}}.
\end{eqnarray}
Using these oscillators, the $R$-matrix is extremely simple and in fact diagonal. Modulo a scalar pre-factor, it reads \cite{Dublin}
\begin{eqnarray}
R = 1 - 2 \eta^\dagger_1 \eta_1.
\end{eqnarray}
Likewise, the factorising twist is also rather simple in terms of free-fermion operators: one can verify that
\begin{eqnarray} 
F_{12}(\theta) = 1 - \mbox{sech}\frac{\theta}{2} \Big[\eta^\dagger_2 + e^{-\frac{\theta}{2}} \eta^\dagger_1 \Big]\eta_1. 
\end{eqnarray}
The labels $1$ and $2$ attached to the $\eta$s have lost the meaning of spin-chain sites, since the canonical transformation mixes the two sites. Therefore, it is not immediately straightforward to compute the opposite twist\footnote{See the discussion around formula (4.48) in \cite{Dublin}.}. One can nevertheless verify that
\begin{eqnarray}
F_{21}^{-1}(-\theta) = 1 + \mbox{csch}\frac{\theta}{2} \Big[\eta^\dagger_2 - e^{\frac{\theta}{2}} \eta^\dagger_1 \Big]\eta_1,
\end{eqnarray}
from which the relation $F_{21}^{-1}(-\theta)F_{12}(\theta) = 1 - 2 \eta^\dagger_1 \eta_1 = R$ follows by virtue of the canonical anti-commutation relations obeyed by the $\eta$s and of known identities of hyperbolic functions. 

\section{Mixed flux bacgkrounds\label{s8}}

We can apply the same strategy to the case of mixed RR and NS-NS flux in the limit where we recover the difference form \cite{gamma2}. The procedure is of course general as it relies on the universal $R$-matrix, but in this limiting case the expressions are more immediately recognisable and easier to handle.

By following the same steps as we did in the massless pure RR case which we have previously discussed, and by implementing the same series of manipulations which are outlined below formula (3.1) in \cite{gamma2}, we obtained the following expression for the factorising twist:
\begin{eqnarray}
F_{12}(q_1,q_2) = \begin{pmatrix}1&0&0&0\\0&1&0&0\\0&-\frac{(e^{\frac{2 i \pi}{k}}-1)\sqrt{q_1q_2}}{q_2 - e^{\frac{2 i \pi}{k}}q_1}&\frac{e^{\frac{i\pi}{k}}(q_1-q_2)}{e^{\frac{2 i \pi}{k}}q_1 - q_2}&0\\0&0&0&\frac{e^{\frac{i\pi}{k}}(q_1-q_2)}{e^{\frac{2 i \pi}{k}}q_1 - q_2}\end{pmatrix},
\end{eqnarray}
$k$ being the mixed-flux parameter. Besides being lower-triangular as expected, this expression is also clearly scale-invariant and of difference form: in fact, by expressing the momenta $q_i$ using the familiar massless (for definiteness right-moving) parametrisation $q_i = e^{\theta_i}$, and setting $\theta = \theta_1 - \theta_2$, one obtains 
\begin{eqnarray}
F_{12}(q_1,q_2) = F_{12}(\theta) = \begin{pmatrix}1&0&0&0\\0&1&0&0\\0&\mbox{csch}(\frac{\pi}{k} -  \frac{\theta}{2}) \, \sinh \frac{\pi}{k}&-\mbox{csch}(\frac{\pi}{k} - \frac{\theta}{2}) \, \sinh \frac{\theta}{2}&0\\0&0&0&-\mbox{csch}(\frac{\pi}{k} - \frac{\theta}{2}) \, \sinh \frac{\theta}{2}\end{pmatrix}.
\end{eqnarray}
By direct calculation one verifies that multiplying $F_{21}^{-1}(q_2,q_1) F_{12}(q_1,q_2)$ exactly reproduces the $R$-matrix in formula (3.3) of \cite{gamma2} (where the ordering $q_2,q_1$ is equivalent to $\theta \to -\theta$).

The level-zero symmetry in this particular case is known \cite{gamma2} to be a deformed (and, in particular, non co-commutative) version of ${\cal{N}}=2$ massless relativistic supersymmetry:
\begin{eqnarray}
\Delta_\rho(Q) = \sqrt{q_1} \, E_{12} \otimes 1_2 + e^{\frac{i \pi}{k}} 1_2 \otimes \sqrt{q_2}\, E_{12}, \qquad \Delta_\rho(Q^\dagger) = \sqrt{q_1} \, E_{21} \otimes 1_2 + e^{\frac{-i \pi}{k}} 1_2 \otimes \sqrt{q_2} \, E_{21}.
\end{eqnarray}
After the action of the twist the $R$-matrix is again equal to the identity, hence the coproducts ought to be co-commutative: we obtain
\begin{eqnarray}
\tilde{\Delta}_\rho(Q) = a(q_1,q_2) \, 1_2\otimes E_{12} + a(q_2,q_1) \, E_{12}\otimes 1_2, \qquad \tilde{\Delta}_\rho(Q^\dagger) = b(q_1,q_2) \, 1_2\otimes E_{12} + b(q_2,q_1) \, E_{12}\otimes 1_2,\label{exo}\nonumber\\
\end{eqnarray}
with 
\begin{eqnarray}
a(q_1,q_2) = \frac{e^{-\frac{i \pi}{k}}\sqrt{q_2}\, (q_1 - e^{-\frac{2i \pi}{k}} q_2)}{q_1-q_2}, \qquad b(q_1,q_2) = e^{-\frac{i \pi}{k}} \sqrt{q_2}.
\end{eqnarray}
The expressions (\ref{exo}) are indeed manifestly co-commutative.

Although this issue was not explored in \cite{gamma2}, we have found the presence of the secret symmetry in this mixed-flux situation as well. There is a level-zero version:
\begin{eqnarray}
\Delta_\rho(b_0) = (E_{11}-E_{22}) \otimes 1_2 + 1_2 \otimes (E_{11}-E_{22}),
\end{eqnarray}
with $F$ the fermionic number; there is also as usual a level-one version, which is of a comparable level of complication as the massive version:
\begin{eqnarray}
&&\Delta_\rho(b_1) = i \, \mbox{cot} \frac{i \pi}{k} \Big[q_1 \, (E_{11}-E_{22}) \otimes 1_2 + 1_2 \otimes q_2 \, (E_{11}-E_{22}) \Big]\qquad \qquad \qquad \nonumber\\
&&\qquad \qquad \qquad \qquad \qquad \qquad+ 2 \sqrt{q_1 q_2} \, e^{\frac{i \pi}{k}} E_{21}\otimes E_{12} + 2 \sqrt{q_1 q_2} \, e^{\frac{-i \pi}{k}} E_{12}\otimes E_{21}. 
\end{eqnarray} 
Notice that one always has to remember that the relative sign in the tail of the secret symmetry must be ``plus'', since a minus would mean (considering the fermionic statistics) that we are just finding the symmetry $\Delta_\rho(Q) . \Delta_\rho(Q^\dagger)$, which simply descends from the fact that $\Delta_\rho(Q)$ and $\Delta_\rho(Q^\dagger)$ are separately symmetries and therefore is not independent.

After the twist, the level zero ``secret'' symmetry becomes the same as the twisted RR level zero secret symmetry, while the level one version become something rather complicated which is not very illuminating, except for the fact that it is verifiably co-commutative. 

\subsection{Zhukovsky variables}

It finally is worth reporting for completeness the representation of the factorising twist in terms of Zhukovsky variables, which applies to massive particles as well. With a particular choice of the so-called {\it frame factors}, one has
\begin{eqnarray}
F_{12}(x_1^\pm,x_2^\pm) = \begin{pmatrix}1&0&0&0\\0&1&0&0\\0&\frac{i \eta_1 \eta_2}{x_1^+ - x_2^-}&\frac{x_1^- - x_2^+}{x_1^+ - x_2^-}&0\\0&0&0&\frac{x_1^- - x_2^+}{x_1^+ - x_2^-}\end{pmatrix}, \qquad \eta_i = \sqrt{i (x_i^--x_i^+)}, \qquad i=1,2.
\end{eqnarray}
Althogh written in a specific representation for the supersymmetry generators, when decorated with appropriate bound state numbers as in \cite{unive} this formula does retain some degree of universality, in the sense that it does not require the mass-shell condition to be derived and therefore it applies for general short representations.

\section{$AdS_2$ and the Universal $R$-matrix\label{s9}}

We shall now discuss the case of the $AdS_2$ integrable system and show that one can obtain the $R$-matrix in a specific limit from the general framework of the universal $R$-matrix of $\alg{gl}(1|1)$ which we have been using for $AdS_3$ as well. This is the first realisation of a quantum (as opposed to classical) $AdS_2$ $R$-matrix along these lines.  

The supersymmetry generators relevant for the $AdS_2$ integrable $R$-matrix in the massless BMN limit (taken with the right-moving chirality for both scattering particles for simplicity) have been studied in \cite{Andrea2} - see also \cite{Hoare:2014kma}, and are given by
\begin{eqnarray}
\check{Q}_i = e^{\frac{\theta_i}{2}-i\frac{\pi}{4}}\sqrt{\frac{1}{2}}\begin{pmatrix}0&1\\1&0\end{pmatrix}, \qquad \check{S}_i = e^{\frac{\theta_i}{2}+i\frac{\pi}{4}}\sqrt{\frac{1}{2}}\begin{pmatrix}0&1\\1&0\end{pmatrix}.\label{rep}
\end{eqnarray}
The notation which we are using in this section for the algebra generators is $\check{Q} \equiv \check{\rho}(q)$, $\check{S} \equiv \check{\rho}(s)$, $P \equiv \check{\rho}(w)$ and $P^\dagger \equiv \check{\rho}(w^\dagger)$, where $\check{\rho}$ denotes the matrix representation, and $x \in \cal{A}'$, while the index $i$ appearing in (\ref{rep}) denotes whether this representation is applied to particle $1$ or $2$ (with rapidity $\theta_1$, resp. $\theta_2$). The algebra $\cal{A}'$ is a defined by
\begin{eqnarray}
\{q,q\} = \frac{w}{2}, \qquad \{s,s\} = \frac{w^\dagger}{2}, \qquad [w,.] = [w^\dagger,.]=0,
\end{eqnarray}
where the dot represents any element in $\cal{A}'$.

We want to apply  - purely empirically as a matter of exercise, and working exclusively with the matrix representation - the same Khoroshkin-Tolstoy formula as in the case of $AdS_3$. This involves computing
\begin{eqnarray}
\frac{\check{Q}\otimes \check{S}}{\lambda_1 - \lambda_2} \label{dis}
\end{eqnarray}
in the representation $\check{\rho}_1 \otimes \check{\rho}_2$: here $\check{\rho}_i$ is also used to denote the representation $\check{\rho}$ taken with parameter $\theta_i$, $i=1,2$ (this being the physical particle rapidity in this particular problem).

The expression (\ref{dis}) turns out to be equal to
\begin{eqnarray}
\frac{\check{S}\otimes \check{Q}}{\lambda_1 - \lambda_2}
\end{eqnarray}
in the representation $\check{\rho}_1 \otimes \check{\rho}_2$. The denominators are as usual the result of (minus) the sum $\sum_{n=0}^\infty \lambda_1^n \, \lambda_2^{-n-1}$ computed from the matrices $\check{Q}_n \equiv \lambda^n \check{Q}, \, \check{S}_n \equiv \lambda^n \check{S}$. The philosophy we are following here is that we would simply insert these matrices in the places where the representation of the Khoroshkin-Tolstoy formula would normally foresee the appearance of $\rho(e_n)$ and $\rho(f_n)$, and see what the calculation gives - purely as a matter of an empirical calculation. We are hoping then to draw some potentially useful intuition from this exercise. Physically we are in the situation of not knowing the algebra and only knowing the $R$-matrix, thereby having to try to gain as much intuition as we can on a possible algebraic reformulation of the physical problem, from such types of empyrical calculations.  

We use the fact that physical considerations require $\lambda_i = e^{\theta_i}$ (up to an irrelevant additive constant). We also send the difference $\theta = \theta_1 - \theta_2 \to \infty$ according to the arguments outlined in \cite{Andrea2}, and obtain
\begin{eqnarray}
\frac{\check{Q}\otimes \check{S}}{\lambda_1 - \lambda_2} = -\frac{1}{2} \mbox{csch}\frac{\theta}{2} \to -  e^{-\frac{\theta}{2}} \qquad \theta \to \infty.
\end{eqnarray}
This means that the exponents of both root factors $R_+$ and $R_-$ in the universal formula of Khoroshkin and Tolstoy's (\ref{pro}) are small, hence we can adopt a linearised approximation
\begin{eqnarray}
R_+ \to 1_2\otimes 1_2 - \frac{\check{Q}\otimes S}{\lambda_1 - \lambda_2}, \qquad  R_- \to1_2\otimes 1_2 + \frac{\check{S}\otimes \check{Q}}{\lambda_1 - \lambda_2}.
\end{eqnarray} 
Note that this would normally be the case without the need of any approximation, since the root elements are supersymmetry generators which normally are nilpotent (as in the case of $AdS_3$). In $AdS_2$, however, this is clearly not the case as $(\check{Q})^2 = P \neq 0$ and $\check{S}^2 = P^\dagger \neq 0$, with $P$ and $P^\dagger$ proportional to the identity matrix, therefore the root factors in the universal formula do not truncate\footnote{The exact resummation would produce trigonometric functions of the evaluation parameter, which we would then expand to the linearised order.}. Nevertheless we can proceed as in the standard case if we set ourselves in the asymptotic limit, which is reminiscent of the analysis developed in \cite{Andrea2} for these $R$-matrices.
 
Always following the philosophy expressed above, the Cartan factors $R_1$ and $R_2$ are computed in the familiar way and are exactly given by (\ref{R1e2}) with $\lambda_i = e^{\theta_i}$ and $a_i d_i = -2 c \,e^{\theta_i}$. We then multiply all the factors $R_+, R_1, R_2, R_-$ together, normalise the entry $|1\rangle\otimes |1\rangle \to |1\rangle \otimes |1\rangle$ to $1$, and take $\theta$ to be large. We get in this way exactly the expression for the $R$-matrix which was dubbed {\it Solution 5} in \cite{Andrea2} for the choice $\alpha=1$ which we have made in (\ref{rep}): 
\begin{eqnarray}
R = \begin{pmatrix}1&0&0&e^{\frac{\theta}{2}}\\0&1&e^{\frac{\theta}{2}}&0\\0&e^{\frac{\theta}{2}}&-1&0\\e^{\frac{\theta}{2}}&0&0&-1\end{pmatrix},\label{5}
\end{eqnarray}
where we can see the typically $8$-vertex form of the $AdS_2$ $R$-matrices. Likewise, in perfect consonance with the arguments of \cite{Andrea2}, by taking the $\theta \to -\infty$ asymptotic limit we obtain was was called {\it Solution 3} in \cite{Andrea2} for the choice $\alpha=1$:
\begin{eqnarray}
R = \begin{pmatrix}1&0&0&-e^{\frac{-\theta}{2}}\\0&-1&e^{\frac{-\theta}{2}}&0\\0&e^{\frac{-\theta}{2}}&1&0\\-e^{\frac{-\theta}{2}}&0&0&-1\end{pmatrix}.\label{3}
\end{eqnarray}
We remind that we are studying worldsheet right-movers, which fits with the form of these $R$-matrices. As in $AdS_3$, it has been necessary to change $\theta \to - \theta$ in the final step of the derivation to match with (\ref{5}) and (\ref{3}). Despite being here obtained in an asymptotic limit, the $R$-matrices (\ref{5}) and (\ref{3}) exactly satisfy the Yang-Baxter equation \cite{Andrea2}.

The procedure of taking an asymptotic limit prevents us from obtaining the factorising twist as we instead did for $AdS_3$. In a sense, it is likely that the factorisation might occurr between a twist for solution $5$ and an inverse opposite twist for solution $3$, which always seem to be paired up - see also \cite{gamma1}. One complicating component in the case of $AdS_2$ is that the entries of the type $E_{12} \otimes E_{12}$ and $E_{21} \otimes E_{21}$ - which are the genuinely $8$-vertex type - remain the same up to a sign under ${}^{op}$. Likewise, under inversion they do remain in the same triangle of the matrix, therefore if one assumes a twist which is lower triangular, even admitting that it has one of the extreme $8$-vertex type entries turned on, there is no way to generate the extreme opposite entry (which the $R$-matrix obviously has) by computing something like $F_{21}^{-1} F_{12}$. 

The only way would be either to have a ``full'' twist (with lower and upper triangular parts) or two twists, as we anticipated above, of the sort $U_{21}^{-1} V_{12}$. Indeed, we have verified that the following decomposition holds:
\begin{eqnarray}
R_{sol.5}(\theta) = U_{21}^{-1}(-\theta) V_{12}(\theta),
\end{eqnarray}     
where
\begin{eqnarray}
V_{12}(\theta) \sim R_2 R_- \qquad \theta \to \infty
\end{eqnarray}
(in the solution-$5$ regime) and
\begin{eqnarray}
U_{12}(\theta) \sim R_+ R_1 \qquad \theta \to -\infty
\end{eqnarray}
(in the solution-$3$ regime), again taking only the first order approximation of the asymptotic expansion\footnote{Specifically, such statements always refer to the expansion in the variable $\xi = e^{\frac{\theta}{2}}$.}. We have also disregarded a global scalar factor in front of the $R$-matrix, since in any case the dressing factor is not a concern of this paper (which is focused on purely algebraic characteristics).

\section{Conclusions}

In this paper we have obtained the factorising twists for the massless $AdS_3$ and $AdS_2$ integrable $R$-matrices, and attempted to transfer to lower-dimensional AdS/CFT the programme pioneered by Maillet et al \cite{Maillet,Maillet2} for ordinary spin-chains. We have derived the factorising twists from the universal $R$-matrix of the $\alg{gl}(1|1)$ Yangian double, and studied the RTT relations for the two- and three-site monodromy matrix. We have shown how the twist can be used to compute a simple scalar product, and we have implemented our construction in the langauge  of free fermions developed in \cite{Dublin}. Finally, we have shown for the first time how to obtain the massless $AdS_2$ quantum $R$-matrix from the Yangian universal $R$-matrix with a particular limiting procedure, and we have computed a peculiar factorising twist for this case as well.  

Even though much of the analysis proceeds as in \cite{Maillet}, ultimately the models which we consider here are harder and we can make much less progress in the technical implementation. It will be very important in future work to try and overcome these obstacles, and obtain a manageable expression for the twist on a generic number of sites of the mondromy matrix, and arbitrary permutations of the sites \cite{Maillet}. General formulas do exist in the literature for ordinary spin-chain, but even without the complications which we have here the general formulas tend to be overwhelmingly complicated. It will also be extremely useful to pursue further the extension to the massive representations, which we have here only shown that it is possible, by using the Zhukovsky variables.

It would be interesting to explore whether the classification of \cite{Marius,NewMarius} can be ``filtered'', in a way, according to which $R$-matrices admit a factorising twist and which perhaps do not. This might complement the analysis of which quantum group structures underlie the various categories of models, and whether such structures do reiterate in any sense the positive-negative-Cartan traingular decomposition which is typical of simple Lie (super)algebras. 

It would be nice to see whether these ideas can be connected to the separation of variables programme \cite{bgood}. Although different, the spirit of finding a change of basis which simplifies the algebraic Bethe ansatz is reminiscent of those advances, likewise the use which is made of these transformations to compute correlation functions. 

It would be very interesting to connect with other approaches to form factors \cite{fofa} and fully exploit the factorisability of the $R$-matrix. It would be nice to take advantage of the fact that the $AdS_5$ $R$-matrix, in all its representations, should naturally admit a factorising twist, since the quantum group it is based upon does respect the triangular structure \cite{AlerevY}, see also the recent \cite{les}. Such twist is probably significantly more complicated, and perhaps practically far less manageable, than the ones which we have found in this paper.   

Drinfeld twists have appeared in recent literature (the most recent to our knowledge being \cite{twists}) to implement deformations, while here we have in a sense undone the twisting to strip the $R$-matrix down to identity. It would be interesting however to connect the two complementary views. 

\section{Acknowledgments}

We thank Juan Miguel Nieto for reading the manuscript and spotting several typos. We thank Patrick Dorey for discussions and for the suggestion to explore the methods developed by Maillet et al to study form factors. We very much thank Suvajit Majumder for discussions about the universal $R$-matrix of the $\alg{gl}(1|1)$ Yangian. We thank Sergey Frolov and Tristan McLoughlin for interesting discussions, and Fabrizio Nieri for an ongoing collaboration on the Lukyanov approach to form factors. We thank the anonymous referee for several very important comments and remarks on the mathematical consistency of the notation used. This work is supported by the EPSRC-SFI grant EP/S020888/1 {\it Solving Spins and Strings}. 

\section*{\label{sec:Data}Data Access Statement}

No data beyond those presented in this paper are needed to validate its results.


\begin{thebibliography}{99}

\bibitem{Bogdan}
A.~Babichenko, B.~Stefa{\'n}ski, and K.~Zarembo,
{\em Integrability and the {$AdS_3/CFT_2$} correspondence,} 
{JHEP {\bf 1003} (2010) 058}
  [\arXivlink{0912.1723}].
  
  \bibitem{Sundin:2012gc}
P.~Sundin and L.~Wulff,
{\em Classical integrability and quantum aspects of the {$AdS_3 \times S^3
  \times S^3 \times S^1$} superstring,} 
{JHEP {\bf 1210} (2012) 109}
  [\arXivlink{1207.5531}].
 
\bibitem{rev3}
  A.~Sfondrini,
  {\em Towards integrability for $AdS_3/CFT_2$,} 
  J.\ Phys.\ A {\bf 48} (2015)  023001
  [\arXivlink{1406.2971}]. 
 
\bibitem{Borsato:2016hud}
  R.~Borsato,
  {\em Integrable strings for AdS/CFT,} 
  [\arXivlink{1605.03173}].
 
\bibitem{Beisertreview}
N.~Beisert, C.~Ahn, L.~F.~Alday, Z.~Bajnok, J.~M.~Drummond, L.~Freyhult, N.~Gromov, R.~A.~Janik, V.~Kazakov and T.~Klose, \textit{et al.}
\emph{Review of AdS/CFT Integrability: An Overview,}
Lett. Math. Phys. \textbf{99} (2012), 3-32
[\arXivlink{1012.3982}].

\bibitem{Foundations}
G.~Arutyunov and S.~Frolov,
\emph{Foundations of the $AdS_5 \times S^5$ Superstring. Part I,}
J. Phys. A \textbf{42} (2009), 254003
[\arXivlink{0901.4937}].

\bibitem{OhlssonSax:2011ms}
O.~Ohlsson Sax and B.~Stefa{\'n}ski,
{\em Integrability, spin-chains, and the {$AdS_3/CFT_2$} correspondence,} 
{JHEP {\bf 1108} (2011) 029}
  [\arXivlink{1106.2558}].

\bibitem{seealso3}
R.~Borsato, O.~Ohlsson Sax, and A.~Sfondrini,
{\em {A dynamic $\mathfrak{su}(1|1)^2$ $S$-matrix for $AdS_3/CFT_2$},} 
{JHEP {\bf 1304} (2013) 113}
  [\arXivlink{1211.5119}].

\bibitem{Borsato:2012ss}
R.~Borsato, O.~Ohlsson Sax, and A.~Sfondrini,
{\em {All-loop Bethe ansatz equations for $AdS_3/CFT_2$},} 
{JHEP {\bf 1304} (2013) 116}
  [\arXivlink{1212.0505}].

\bibitem{Borsato:2013qpa}
R.~Borsato, O.~Ohlsson Sax, A.~Sfondrini, B.~Stefa\'nski, and A.~Torrielli,
{\em {The all-loop integrable spin-chain for strings on AdS$_3 \times S^3
  \times T^4$: the massive sector},} 
{JHEP {\bf 1308} (2013) 043}
  [\arXivlink{1303.5995}].


\bibitem{Borsato:2013hoa}
  R.~Borsato, O.~Ohlsson Sax, A.~Sfondrini, B.~Stefa\'nski, Jr. and A.~Torrielli,
  {\em Dressing phases of $AdS_3/CFT_2$,} 
  Phys.\ Rev.\ D {\bf 88} (2013) 066004
  [\arXivlink{1306.2512}].

\bibitem{Rughoonauth:2012qd}
N.~Rughoonauth, P.~Sundin, and L.~Wulff,
{\em Near {BMN} dynamics of the {$AdS_3 \times S^3 \times S^3 \times S^1$}
  superstring,} 
{JHEP {\bf 1207} (2012) 159}
  [\arXivlink{1204.4742}].

  \bibitem{PerLinus}
P.~Sundin and L.~Wulff,
  {\em The complete one-loop BMN $S$-matrix in $AdS_{3}\times  S^{3}\times T^{4}$,} 
  JHEP {\bf 1606} (2016) 062
[\arXivlink{1605.01632}].

\bibitem{CompleteT4}
  R.~Borsato, O.~Ohlsson Sax, A.~Sfondrini and B.~Stefa\'nski,
  {\em The complete AdS$_{3} \times$ S$^3 \times$ T$^4$ worldsheet S matrix,} 
  JHEP {\bf 1410} (2014) 66
  [\arXivlink{1406.0453}]. 
  
 \bibitem{Borsato:2015mma}
R.~Borsato, O.~Ohlsson Sax, A.~Sfondrini and B.~Stefa\'nski,
  {\em The $AdS_3\times S^3\times S^3 \times S^1$ worldsheet S matrix,} 
  J.\ Phys.\ A {\bf 48} (2015)  415401
  [\arXivlink{1506.00218}].
  
\bibitem{Beccaria:2012kb}
M.~Beccaria, F.~Levkovich-Maslyuk, G.~Macorini, and A.~Tseytlin,
{\em {Quantum corrections to spinning superstrings in $AdS_3\times S^3 \times
  M^4$: determining the dressing phase},} 
{JHEP {\bf 1304} (2013) 006}
  [\arXivlink{1211.6090}].

\bibitem{Sundin:2013ypa}
P.~Sundin and L.~Wulff,
{\em {World-sheet scattering in $AdS_3/CFT_2$},} 
{JHEP {\bf 1307} (2013) 007}
  [\arXivlink{1302.5349}].

\bibitem{Bianchi:2013nra}
L.~Bianchi, V.~Forini, and B.~Hoare,
{\em {Two-dimensional $S$-matrices from unitarity cuts},} 
{JHEP {\bf 1307} (2013) 088}
  [\arXivlink{1304.1798}].

\bibitem{Bianchi:2013nra1}
O.~T.~Engelund, R.~W.~McKeown and R.~Roiban,
{\em Generalised unitarity and the worldsheet $S$-matrix in $AdS_n \times S^n \times M^{10-2n}$,} 
JHEP {\bf 1308} (2013) 023
[\arXivlink{1304.4281}].

\bibitem{Bianchi:2013nra2}
  L.~Bianchi and B.~Hoare,
  {\em $AdS_3 \times S^3 \times M^4$ string $S$-matrices from unitarity cuts,} 
  JHEP {\bf 1408} (2014) 097
  [\arXivlink{1405.7947}].

 \bibitem{Sax:2012jv}
O.~Ohlsson Sax, B.~Stefa\'nski, and A.~Torrielli,
{\em {On the massless modes of the $AdS_3/CFT_2$ integrable systems},} 
{JHEP {\bf 1303} (2013) 109}
  [\arXivlink{1211.1952}].

  \bibitem{Borsato:2016xns}
  R.~Borsato, O.~Ohlsson Sax, A.~Sfondrini, B.~Stefa\'nski, Jr. and A.~Torrielli,
  \emph{On the dressing factors, Bethe equations and Yangian symmetry of strings on $AdS_3 \times S^3 \times T^4$,} 
  J.\ Phys.\ A {\bf 50} (2017) 024004
  [\arXivlink{1607.00914}].
  
   \bibitem{Sax:2014mea}
  O.~Ohlsson Sax, A.~Sfondrini and B.~Stefa\'nski,
  {\em Integrability and the Conformal Field Theory of the Higgs branch,} 
  JHEP {\bf 1506} (2015) 103
  [\arXivlink{1411.3676}].
 
 \bibitem{Baggio:2017kza}
  M.~Baggio, O.~Ohlsson Sax, A.~Sfondrini, B.~Stefa\'nski and A.~Torrielli,
  {\em Protected string spectrum in AdS3/CFT2 from worldsheet integrability,} 
  [\arXivlink{1701.03501}].
  S.~Majumder, O.~O.~Sax, B.~Stefa\'nski and A.~Torrielli,
\emph{Protected states in $AdS_3$ backgrounds from integrability,}
J. Phys. A \textbf{54} (2021) no.41, 415401
[\arXivlink{2103.16972}].

 \bibitem{Zamol2}
A.~B.~Zamolodchikov and A.~B.~Zamolodchikov,
	{\em Massless factorized scattering and sigma models with topological terms,}
Nucl.\ Phys.\ B {\bf 379} (1992) 602.
P.~Fendley, H.~Saleur and A.~B.~Zamolodchikov,
\emph{Massless flows, 2. The Exact S-matrix approach,}
Int.\ J.\ Mod.\ Phys.\ A {\bf 8} (1993) 5751
[\arXivlink{hep-th/9304051}].
P.~Fendley and K.~A.~Intriligator,
\emph{Exact $N=2$ Landau-Ginzburg flows,}
Nucl. Phys. B \textbf{413} (1994), 653-674
[arXiv:hep-th/9307166 [hep-th]].

\bibitem{Fendley:1993jh}
  P.~Fendley and H.~Saleur,
  \emph{Massless integrable quantum field theories and massless scattering in (1+1)-dimensions,}
  [\arXivlink{hep-th/9310058}].


\bibitem{DiegoBogdanAle}
  D.~Bombardelli, B.~Stefa\'nski and A.~Torrielli,
  \emph{The low-energy limit of $AdS_3/CFT_2$ and its TBA,} 
  JHEP {\bf 1810} (2018) 177
  [\arXivlink{1807.07775}].
 
    \bibitem{Lloyd:2013wza}
T.~Lloyd and B.~Stefa\'nski,
{\em {$AdS_3/CFT_2$, finite-gap equations and massless modes},} 
{JHEP {\bf 1404} (2014) 179}
  [\arXivlink{1312.3268}].

 
 
 \bibitem{Abbott:2012dd}
M.~C.~Abbott,
{\em {Comment on strings in $AdS_3 \times S^3 \times S^3 \times S^1$ at one
  loop},} 
{JHEP {\bf 1302} (2013) 102}
  [\arXivlink{1211.5587}].

  \bibitem{Abbott:2014rca}
  M.~C.~Abbott and I.~Aniceto,
  {\em Macroscopic (and Microscopic) Massless Modes,} 
  Nucl.\ Phys.\ B {\bf 894} (2015) 75
  [\arXivlink{1412.6380}].

  
  \bibitem{MI}
  M.~C.~Abbott and I.~Aniceto,
  \emph{Massless L\"uscher terms and the limitations of the AdS$_3$ asymptotic Bethe ansatz,}
  Phys.\ Rev.\ D {\bf 93} (2016) no.10,  106006
  [\arXivlink{1512.08761}].

\bibitem{Abbott:2020jaa}
  M.~C.~Abbott and I.~Aniceto,
  \emph{Integrable Field Theories with an Interacting Massless Sector,}
 [\arXivlink{2002.12060}].

   \bibitem{Eberhardt:2017fsi}
  L.~Eberhardt, M.~R.~Gaberdiel, R.~Gopakumar and W.~Li,
  {\em BPS spectrum on AdS$_3\times $S$^3 \times $S$^3 \times $S$^1$,} 
  [\arXivlink{1701.03552}]. 
  
\bibitem{Gaber1}
M.~R.~Gaberdiel, R.~Gopakumar and C.~Hull,
  \emph{Stringy AdS$_{3}$ from the worldsheet,} 
  JHEP {\bf 1707} (2017) 090
    [\arXivlink{1704.08665}].

\bibitem{Gaber2}
  L.~Eberhardt, M.~R.~Gaberdiel and W.~Li,
  \emph{A holographic dual for string theory on AdS$_3\times $S$^3 \times $S$^3 \times $S$^1$,} 
  JHEP {\bf 1708} (2017) 111
  [\arXivlink{1707.02705}].

\bibitem{Gaber3}
  O.~Ohlsson Sax and B.~Stefa\'nski,
  \emph{Closed strings and moduli in AdS$_{3}$/CFT$_{2}$,} 
  JHEP {\bf 1805} (2018) 101
  [\arXivlink{1804.02023}].
  
  \bibitem{Gaber4}
  A.~Dei, M.~R.~Gaberdiel and A.~Sfondrini,
  \emph{The plane-wave limit of ${\rm AdS}_3 \times {\rm S}^3 \times {\rm S}^3 \times {\rm S}^1$,} 
  JHEP {\bf 1808} (2018) 097
  [\arXivlink{1805.09154}].

\bibitem{Gaber5}
  A.~Dei and A.~Sfondrini,
  \emph{Integrable spin chain for stringy Wess-Zumino-Witten models,} 
  JHEP {\bf 1807} (2018) 109
  [\arXivlink{1806.00422}].


\bibitem{GaberdielUltimo}
A.~Dei, L.~Eberhardt and M.~R.~Gaberdiel,
  {\em Three-point functions in AdS$_3$/CFT$_2$ holography,}
  [\arXivlink{1907.13144}].
A.~Dei, M.~R.~Gaberdiel, R.~Gopakumar and B.~Knighton,
\emph{Free field world-sheet correlators for $AdS_3$,}
JHEP \textbf{02} (2021), 081
[\arXivlink{2009.11306}].
  M.~R.~Gaberdiel, B.~Knighton and J.~Vo\v{s}mera,
\emph{D-branes in $AdS_3 \times S_2 \times T^4$ at $k = 1$ and their holographic duals,}
JHEP \textbf{12} (2021), 149
[\arXivlink{2110.05509}].
  
\bibitem{Prin}
A.~Prinsloo,
  {\em D1 and D5-brane giant gravitons on $AdS_3 \times S^3 \times S^3 \times S^1$,} 
  JHEP {\bf 1412} (2014) 094
  [\arXivlink{1406.6134}].

\bibitem{Prin1}
A.~Prinsloo, V.~Regelskis and A.~Torrielli,
  {\em Integrable open spin-chains in $AdS_3/CFT_2$ correspondences,} 
  Phys.\ Rev.\ D {\bf 92} (2015) no.10,  106006
  [\arXivlink{1505.06767}].
  

\bibitem{Abbott:2015mla}
M. C. Abbott, J. Murugan, S. Penati, A. Pittelli, D. Sorokin, P. Sundin, J. Tarrant, M. Wolf and L. Wulff,
  {\em T-duality of Green-Schwarz superstrings on 
  $AdS_d \times S^d \times M^{10-2d}$,} 
  JHEP {\bf 1512} (2015) 104
  [\arXivlink{1509.07678}].

\bibitem{Per9}
  L.~Wulff,
  {\em On integrability of strings on symmetric spaces,} 
  JHEP {\bf 1509} (2015) 115
  [\arXivlink{1505.03525}].

\bibitem{Hoare:2018jim}
  B.~Hoare, N.~Levine and A.~A.~Tseytlin,
  {\em On the massless tree-level $S$-matrix in 2d sigma models,} 
  J.\ Phys.\ A {\bf 52} (2019) no.14,  144005
  [\arXivlink{1812.02549}].

\bibitem{AntonioMartin}
  A.~Pittelli, A.~Torrielli and M.~Wolf,
  {\em Secret symmetries of type IIB superstring theory on $AdS_3 \times S^3 \times M^4$,} 
  J.\ Phys.\ A {\bf 47} (2014) no.45,  455402
  [\arXivlink{1406.2840}].



\bibitem{Regelskis:2015xxa}
  V.~Regelskis,
  {\em Yangian of $AdS_3/CFT_2$ and its deformation,} 
  J.\ Geom.\ Phys.\  {\bf 106} (2016) 213
  [\arXivlink{1503.03799}].

 
  

\bibitem{olobo}
O.~Ohlsson Sax and B.~Stefa\'nski,
\emph{On the singularities of the RR $AdS_3 \times S^3 \times T^4$ S matrix,}
J. Phys. A \textbf{53} (2020) no.15, 155402
[\arXivlink{1912.04320}].

\bibitem{Baggio}
 M.~Baggio and A.~Sfondrini,
  \emph{Strings on NS-NS backgrounds as integrable deformations,}
  Phys. \ Rev. \ D {\bf 98} (2018) 021902
   [\arXivlink{1804.01998}].
{}
  A.~Dei and A.~Sfondrini,
  \emph{Integrable spin chain for stringy Wess-Zumino-Witten models,}
  JHEP {\bf 1807} (2018) 109
  [\arXivlink{1806.00422}].
{}
  B.~Hoare and A.~A.~Tseytlin,
  \emph{Towards the quantum S-matrix of the Pohlmeyer reduced version of $AdS_5 \times S^5$ superstring theory,}
  Nucl.\ Phys.\ B {\bf 851} (2011) 161
  [\arXivlink{1104.2423}].
{}
  B.~Hoare,
  \emph{Towards a two-parameter q-deformation of AdS$_3 \times S^3 \times M^4$ superstrings,}
  Nucl.\ Phys.\ B {\bf 891} (2015) 259
  [\arXivlink{1411.1266}].
{}
  G.~Giribet, C.~Hull, M.~Kleban, M.~Porrati and E.~Rabinovici,
  \emph{Superstrings on AdS$_{3}$ at $k =1$,}
  JHEP {\bf 1808} (2018) 204
  [\arXivlink{1803.04420}].
{}
  M.~R.~Gaberdiel and R.~Gopakumar,
  \emph{Tensionless string spectra on AdS$_{3}$,}
  JHEP {\bf 1805} (2018) 085
  [\arXivlink{1803.04423}].
{}
  L.~Eberhardt, M.~R.~Gaberdiel and R.~Gopakumar,
  \emph{The Worldsheet Dual of the Symmetric Product CFT,}
  JHEP {\bf 1904} (2019) 103
  [\arXivlink{1812.01007}].
  A.~Edery,
\emph{Non-singular vortices with positive mass in $2+1$ dimensional Einstein gravity with $AdS_3$ and Minkowski background,}
JHEP \textbf{01} (2021) 166
[\arXivlink{2004.09295}].
  
\bibitem{JuanMiguelAle}
J.~M.~Nieto Garc\'ia and A.~Torrielli,
{\em Norms and scalar products for $AdS_3$,}
J. \ Phys. \ A {\bf 53} (2020) 145401
[\arXivlink{1911.06590}].


 \bibitem{QSC}
A.~Cavagli\`a, N.~Gromov, B.~Stefa\'nski, Jr., Jr. and A.~Torrielli,
\emph{Quantum Spectral Curve for $AdS_3/CFT_2$: a proposal,}
JHEP \textbf{12} (2021), 048
[\arXivlink{2109.05500}].
S.~Ekhammar and D.~Volin,
\emph{Monodromy Bootstrap for $\alg{su}(2|2)$ Quantum Spectral Curves: From Hubbard model to $AdS_3/CFT_2$,}
[\arXivlink{2109.06164}].
  
\bibitem{AleSSergey}
S.~Frolov and A.~Sfondrini,
\emph{Massless S matrices for $AdS_3/CFT_2$,}
[\arXivlink{2112.08895}].
S.~Frolov and A.~Sfondrini,
\emph{New Dressing Factors for $AdS_3/CFT_2$,}
[\arXivlink{2112.08896}].
S.~Frolov and A.~Sfondrini,
\emph{Mirror Thermodynamic Bethe Ansatz for $AdS_3/CFT_2$,}
[\arXivlink{2112.08898}].

\bibitem{Seibold:2022mgg}
F.~K.~Seibold and A.~Sfondrini,
\emph{Transfer matrices for $AdS_3/CFT_2$,}
[\arXivlink{2202.11058}].

 \bibitem{Hopf}
C.~Gomez and R.~Hernandez,
\emph{The Magnon kinematics of the AdS/CFT correspondence,}
JHEP \textbf{11} (2006), 021
[\arXivlink{hep-th/0608029}]. 
J.~Plefka, F.~Spill and A.~Torrielli,
\emph{On the Hopf algebra structure of the AdS/CFT S-matrix,}
Phys. Rev. D \textbf{74} (2006), 066008
[\arXivlink{hep-th/0608038}].
  
  
  \bibitem{CesarRafa}
  C.~Gomez and R.~Hern\'andez,
  {\em Quantum deformed magnon kinematics,}  
JHEP {\bf 0703} (2007) 108
  [\arXivlink{hep-th/0701200}].

\bibitem{Charles}
  C.~A.~S.~Young,
  {\em $q$-deformed supersymmetry and dynamic magnon representations,}
  J.\ Phys.\ A {\bf 40} (2007) 9165
  [\arXivlink{0704.2069}].


\bibitem{Riccardo}
  R.~Borsato and A.~Torrielli,
  {\em $q$ -Poincar\'e supersymmetry in AdS$_5$ / CFT$_4$,}
  Nucl.\ Phys.\ B {\bf 928} (2018) 321
  [\arXivlink{1706.10265}].
  
  \bibitem{qseealso}
  B.~Hoare, T.~J.~Hollowood and J.~L.~Miramontes,
  \emph{A Relativistic Relative of the Magnon S-Matrix,}
  JHEP {\bf 1111} (2011) 048
  [\arXivlink{1107.0628}].
  B.~Hoare, T.~J.~Hollowood and J.~L.~Miramontes,
  \emph{q-Deformation of the $AdS_5 \times S^5$ Superstring S-matrix and its Relativistic Limit,}
  JHEP {\bf 1203} (2012) 015
  [\arXivlink{1112.4485}].
{}
  

\bibitem{JoakimAle}
  J.~Str\"omwall and A.~Torrielli,
 \emph{AdS$_{3}$/CFT$_{2}$ and $q$-Poincar\'e superalgebras,}
  J.\ Phys.\ A {\bf 49} (2016) no.43,  435402
  [\arXivlink{1606.02217}].
  


\bibitem{BorStromTorri}
  R.~Borsato, J.~Str\"omwall and A.~Torrielli,
 \emph{ $q$-Poincar\'e invariance of the AdS$_3$/CFT$_2$ $R$-matrix,}
  Phys.\ Rev.\ D {\bf 97} (2018) no.6,  066001
  [\arXivlink{1711.02446}].
  

\bibitem{Andrea}
  A.~Fontanella and A.~Torrielli,
  \emph{Massless sector of AdS$_3$ superstrings: A geometric interpretation,}
  Phys.\ Rev.\ D {\bf 94} (2016) no.6,  066008
  [\arXivlink{1608.01631}].
  J.~M.~Nieto Garc\'\i{}a, A.~Torrielli and L.~Wyss,
\emph{Boost generator in $AdS_3$ integrable superstrings for general braiding,}
JHEP \textbf{07} (2020), 223
[\arXivlink{2004.02531}].
  J.~M.~Nieto Garc\'\i{}a, A.~Torrielli and L.~Wyss,
\emph{Boosts superalgebras based on centrally-extended $\alg{su}(1|1)^2$,}
J. Geom. Phys. \textbf{164} (2021), 104172
[\arXivlink{2009.11171}].
  

\bibitem{gamma1}
  A.~Fontanella and A.~Torrielli,
  \emph{Geometry of Massless Scattering in Integrable Superstring,}
  JHEP {\bf 1906} (2019) 116
  [\arXivlink{1903.10759}].
  
\bibitem{gamma2}
  A.~Fontanella, O.~Ohlsson Sax, B.~Stefa\'nski and A.~Torrielli,
  \emph{The effectiveness of relativistic invariance in AdS$_{3}$,} 
  JHEP {\bf 1907} (2019) 105
  [\arXivlink{1905.00757}].
    
\bibitem{Cagnazzo:2012se}
  A.~Cagnazzo and K.~Zarembo,
  \emph{B-field in $AdS_3/CFT_2$ Correspondence and Integrability,}
  JHEP {\bf 1211} (2012) 133
  [\arXivlink{1209.4049}].
  

\bibitem{s1}
  B.~Hoare and A.~A.~Tseytlin,
  \emph{On string theory on $AdS_3 \times S^3 \times T^4$ with mixed 3-form flux: tree-level S-matrix,}
  Nucl.\ Phys.\ B {\bf 873} (2013) 682
  [\arXivlink{1303.1037}].


   
\bibitem{s2}
  B.~Hoare and A.~A.~Tseytlin,
  \emph{Massive S-matrix of $AdS_3 \times S^3 \times T^4$ superstring theory with mixed 3-form flux,}
  Nucl.\ Phys.\ B {\bf 873} (2013) 395
  [\arXivlink{1304.4099}].


  \bibitem{Babichenko:2014yaa}
  A.~Babichenko, A.~Dekel and O.~Ohlsson Sax,
  \emph{Finite-gap equations for strings on AdS$_{3}$ x S$^{3}$ x T$^{4}$ with mixed 3-form flux,}
  JHEP {\bf 1411} (2014) 122
  [\arXivlink{1405.6087}].


 
\bibitem{seealso12}
  A.~Pittelli,
  \emph{Yangian Symmetry of String Theory on $AdS_3 \times S^3 \times S^3 \times S^1$ with Mixed 3-form Flux,}
  Nucl.\ Phys.\ B {\bf 935} (2018) 271
  [\arXivlink{1711.02468}].


\bibitem{ArkadyBenStepanchuk} 
  B.~Hoare, A.~Stepanchuk and A.~A.~Tseytlin,
  \emph{Giant magnon solution and dispersion relation in string theory in $AdS_3 \times S^3 \times T^4$ with mixed flux,}
  Nucl.\ Phys.\ B {\bf 879} (2014) 318
  [\arXivlink{1311.1794}].
  

  
\bibitem{Lloyd:2014bsa}
  T.~Lloyd, O.~Ohlsson Sax, A.~Sfondrini and B.~Stefa\'nski, jr.,
  \emph{The complete worldsheet S matrix of superstrings on $AdS_3 \times S^3 \times T^4$ with mixed three-form flux,}
  Nucl.\ Phys.\ B {\bf 891} (2015) 570
  [\arXivlink{1410.0866}].
  
  
   \bibitem{OhlssonSax:2018hgc}
  O.~Ohlsson Sax and B.~Stefa\'nski, jr.,
  \emph{Closed strings and moduli in AdS$_{3}$/CFT$_{2}$,}
  JHEP {\bf 1805} (2018) 101
  [\arXivlink{1804.02023}].


  
  \bibitem{ads2}
I.~R.~Klebanov and A.~A.~Tseytlin,
\emph{Intersecting M-branes as four-dimensional black holes,}
Nucl.\ Phys.\ B {\bf 475} (1996) 179
[\arXivlink{hep-th/9604166}].
A.~A.~Tseytlin,
\emph{Harmonic superpositions of M-branes,}
Nucl.\ Phys.\ B {\bf 475} (1996) 149
[\arXivlink{hep-th/9604035}].
M.~J.~Duff, H.~Lu and C.~N.~Pope,
\emph{$AdS_5 \times S^5$ untwisted,}
Nucl.\ Phys.\ B {\bf 532} (1998) 181
[\arXivlink{hep-th/9803061}].
H.~J.~Boonstra, B.~Peeters and K.~Skenderis,
\emph{Brane intersections, anti-de Sitter space-times and dual superconformal theories,}
Nucl.\ Phys.\ B {\bf 533} (1998) 127
[\arXivlink{hep-th/9803231}].
J.~Lee and S.~Lee,
\emph{Mass spectrum of D=11 supergravity on $AdS_2 \times S^2 \times T^7$,}
Nucl.\ Phys.\ B {\bf 563} (1999) 125
[\arXivlink{hep-th/9906105}].

\bibitem{dual}
A.~Strominger,
\emph{$AdS_2$ quantum gravity and string theory,}
JHEP {\bf 9901} (1999) 007
[\arXivlink{hep-th/9809027}].
G.~W.~Gibbons and P.~K.~Townsend,
\emph{Black holes and Calogero models,}
Phys.\ Lett.\ B {\bf 454} (1999) 187
[\arXivlink{hep-th/9812034}].
J.~M.~Maldacena, J.~Michelson and A.~Strominger,
\emph{Anti-de Sitter fragmentation,}
JHEP {\bf 9902} (1999) 011
[\arXivlink{hep-th/9812073}].
C.~Chamon, R.~Jackiw, S.-Y.~Pi and L.~Santos,
\emph{Conformal quantum mechanics as the $CFT_1$ dual to $AdS_2$,}
Phys.\ Lett.\ B {\bf 701} (2011) 503
[\arXivlink{1106.0726}].

\bibitem{gen}
  A.~Castro, D.~Grumiller, F.~Larsen and R.~McNees,
  \emph{Holographic Description of $AdS_2$ Black Holes,}
  JHEP {\bf 0811} (2008) 052
  [\arXivlink{0809.4264}].
{}
D.~Ridout and J.~Teschner,
  \emph{Integrability of a family of quantum field theories related to sigma models,}
  Nucl.\ Phys.\ B {\bf 853} (2011) 327
  [\arXivlink{1102.5716}].
{}
A.~Dabholkar, J.~Gomes and S.~Murthy,
  \emph{Quantum black holes, localization and the topological string,}
  JHEP {\bf 1106} (2011) 019
  [\arXivlink{1012.0265}].
  {}
D.~M.~Hofman and A.~Strominger,
  \emph{Chiral Scale and Conformal Invariance in 2D Quantum Field Theory,}
  Phys.\ Rev.\ Lett.\  {\bf 107} (2011) 161601
  [\arXivlink{1107.2917}].
  {}  
  A.~Almheiri and J.~Polchinski,
\emph{Models of $AdS_2$ Backreaction and Holography,}
[\arXivlink{1402.6334}].
M.~Heinze, B.~Hoare, G.~Jorjadze and L.~Megrelidze,
\emph{Orbit method quantization of the $AdS_2$ superparticle,}
J.\ Phys.\ A {\bf 48} (2015) 31, 315403
[\arXivlink{1504.04175}].
O.~Lunin,
\emph{Bubbling geometries for $AdS_2 \times S^2$,}
[\arXivlink{1507.06670}].
O.~Lechtenfeld and S.~Nampuri,
\emph{A Calogero formulation for four-dimensional black-hole micro states,}
[\arXivlink{1509.03256}].
{}
R.~Borsato, A.~A.~Tseytlin and L.~Wulff,
  \emph{Supergravity background of $\lambda$-deformed model for $AdS_2 \times  S^2$ supercoset,}
  Nucl.\ Phys.\ B {\bf 905} (2016) 264
  [\arXivlink{1601.08192}].
{}
M.~Mezei, S.~S.~Pufu and Y.~Wang,
  \emph{A 2d/1d Holographic Duality,}
  [\arXivlink{1703.08749}].
  {}
S.~Giombi, R.~Roiban and A.~A.~Tseytlin,
  \emph{Half-BPS Wilson loop and $AdS_2/CFT_1$,}
  [\arXivlink{1706.00756}].
{}
J.-G.~Zhou,
\emph{Super 0-brane and GS superstring actions on $AdS_2 \times S^2$,}
Nucl.\ Phys.\ B {\bf 559} (1999) 92
[\arXivlink{hep-th/9906013}].
N.~Berkovits, M.~Bershadsky, T.~Hauer, S.~Zhukov and B.~Zwiebach,
\emph{Superstring theory on $AdS_2 \times S^2$ as a coset supermanifold,}
Nucl.\ Phys.\ B {\bf 567} (2000) 61
[\arXivlink{hep-th/9907200}].

\bibitem{Sorokin:2011rr}
D.~Sorokin, A.~Tseytlin, L.~Wulff and K.~Zarembo,
\emph{Superstrings in $AdS_2 \times S^2 \times T^6$,}
J.\ Phys.\ A {\bf 44} (2011) 275401
[\arXivlink{1104.1793}].

\bibitem{Cagnazzo:2011at}
A.~Cagnazzo, D.~Sorokin and L.~Wulff,
\emph{More on integrable structures of superstrings in $AdS_4 \times \mathbb{C}\mbox{P}^3$ and $AdS_2 \times S^2 \times T^6$ superbackgrounds,}
JHEP {\bf 1201} (2012) 004
[\arXivlink{1111.4197}].

\bibitem{amsw}
J.~Murugan, P.~Sundin and L.~Wulff,
\emph{Classical and quantum integrability in $AdS_2/CFT_1$,}
JHEP {\bf 1301} (2013) 047
[\arXivlink{1209.6062}].
M.~C.~Abbott, J.~Murugan, P.~Sundin and L.~Wulff,
\emph{Scattering in $AdS_2/CFT_1$ and the BES Phase,}
JHEP {\bf 1310} (2013) 066
[\arXivlink{1308.1370}].

\bibitem{Hoare:2014kma}
B.~Hoare, A.~Pittelli and A.~Torrielli,
\emph{Integrable S-matrices, massive and massless modes and the $AdS_2 \times S^2$ superstring,}
JHEP {\bf 1411} (2014) 051
[\arXivlink{1407.0303}].



\bibitem{Arutyunov:2009pw}
G.~Arutyunov, M.~de Leeuw and A.~Torrielli,
\emph{On Yangian and Long Representations of the Centrally Extended $\mathfrak{su}(2|2)$ Superalgebra,}
JHEP {\bf 1006} (2010) 033
[\arXivlink{0912.0209}].

\bibitem{Hoare:2014kmaa}
B.~Hoare, A.~Pittelli and A.~Torrielli,
  \textit{$S$-matrix algebra of the $AdS_2 \times S^2$ superstring,
  Phys.\ Rev.}\ D {\bf 93} (2016)  066006
  [\arXivlink{1509.07587}].
  
\bibitem{Per6}
  R.~Roiban, P.~Sundin, A.~Tseytlin and L.~Wulff,
  {\em The one-loop worldsheet $S$-matrix for the $AdS_n \times S^n \times T^{10-2n}$ superstring,}
  JHEP {\bf 1408} (2014) 160
  [\arXivlink{1407.7883}].

\bibitem{Per10}
  P.~Sundin and L.~Wulff,
  {\em The $AdS_{n} \times S^{n} \times T^{10-2n}$ BMN string at two loops,}
  JHEP {\bf 1511} (2015) 154
  [\arXivlink{1508.04313}].

\bibitem{Andrea2}  
  A.~Fontanella and A.~Torrielli,
   \textit{Massless $AdS_2$ scattering and Bethe ansatz,}
 JHEP {\bf 1709} (2017) 075
[\arXivlink{1706.02634}]. 
  
 
  
  
   \bibitem{Fendley:1990cy}
  P.~Fendley,
  \textit{A Second supersymmetric S-matrix for the perturbed tricritical Ising model,}
  Phys.\ Lett. \ B {\bf 250} (1990) 96.
  

\bibitem{Baxter:1972hz}
  R.~J.~Baxter,
  \textit{Partition function of the eight vertex lattice model,}
  Annals Phys.\  {\bf 70} (1972) 193.

\bibitem{Baxter:1972hz1}
  R.~J.~Baxter,
  \textit{One-dimensional anisotropic Heisenberg chain,}
  Annals Phys. \  {\bf 70} (1972) 323.

  
\bibitem{Schoutens}
  K.~Schoutens,
  \textit{Supersymmetry and Factorizable Scattering,}
  Nucl.\ Phys.\ B {\bf 344} (1990) 665.
  
\bibitem{MC}
  M.~Moriconi and K.~Schoutens,
  \textit{Thermodynamic Bethe ansatz for ${\cal{N}}=1$ supersymmetric theories,}
  Nucl.\ Phys.\ B {\bf 464} (1996) 472
  [\arXivlink{hep-th/9511008}].


\bibitem{Levkovich-Maslyuk:2016kfv}
  F.~Levkovich-Maslyuk,
  \textit{The Bethe ansatz,}
  J.\ Phys.\ A {\bf 49} (2016) 323004
  [\arXivlink{1606.02950}].

\bibitem{Nepo}
L.~A.~Takhtajan and L.~D.~Faddeev,
  \textit{The Quantum method of the inverse problem and the Heisenberg XYZ model,}
  Russ.\ Math.\ Surveys {\bf 34} (1979) 11
   [Usp.\ Mat.\ Nauk {\bf 34} (1979) 13].

\bibitem{Nepo1}
D.~Fioravanti and M.~Rossi,
\textit{From the braided to the usual Yang-Baxter relation,}
  J.\ Phys.\ A {\bf 34} (2001) L567
  [\arXivlink{0107050}].

\bibitem{Nepo2}
  J.~Cao, W.~L.~Yang, K.~Shi and Y.~Wang,
  \textit{Off-diagonal Bethe ansatz solution of the XXX spin-chain with arbitrary boundary conditions,}
  Nucl.\ Phys.\ B {\bf 875} (2013) 152
  [\arXivlink{1306.1742}].

\bibitem{Nepo3}
S.~Belliard and N.~Cramp\'e,
  \textit{Heisenberg XXX Model with General Boundaries: Eigenvectors from Algebraic Bethe Ansatz,}
  SIGMA {\bf 9} (2013) 072
  [\arXivlink{1309.6165}].

\bibitem{Nepo4}
   X.~Zhang, J.~Cao, S.~Cui, R.~I.~Nepomechie, W.~L.~Yang, K.~Shi and Y.~Wang,
  \textit{Bethe ansatz for an AdS/CFT open spin chain with non-diagonal boundaries,}
  JHEP {\bf 1510} (2015) 133
  [\arXivlink{1507.08866}].

\bibitem{Nepo5}
Y.~Wang, W.~L.~Yang, J.~Cao and K.~Shi,
\textit{Off-Diagonal Bethe Ansatz for Exactly Solvable Models,}
Springer, 2015.

\bibitem{Nepo6}
M.~Guica, F.~Levkovich-Maslyuk and K.~Zarembo,
  \textit{Integrability in dipole-deformed N=4 super Yang-Mills,}
  [\arXivlink{1706.07957}].  
   L.~D.~Faddeev and O.~Tirkkonen,
  \textit{Connections of the Liouville model and XXZ spin chain,}
  Nucl.\ Phys.\ B {\bf 453} (1995) 647
  [\arXivlink{hep-th/9506023}].
  D.~Fioravanti and M.~Rossi,
  \textit{A Braided Yang-Baxter algebra in a theory of two coupled lattice quantum KdV: Algebraic properties and ABA representations,}
  J.\ Phys.\ A {\bf 35} (2002) 3647
  [\arXivlink{hep-th/0104002}].
  
 \bibitem{Ale}
A.~Torrielli,
  \textit{On $AdS_2/CFT_1$ transfer matrices, Bethe ansatz and scale invariance,}
  J.\ Phys.\ A {\bf 51} (2018) no.1,  015402
  [\arXivlink{1708.09598}].
  
\bibitem{Ahn}
  C.~Ahn,
  \textit{Thermodynamics and form-factors of supersymmetric integrable field theories,}
  Nucl.\ Phys.\ B {\bf 422} (1994) 449
  [{}\arXivlink{hep-th/9306146}].

\bibitem{Zamolodchikov:1991vh}
A.~B.~Zamolodchikov,
\textit{Thermodynamic Bethe ansatz for RSOS scattering theories,}
Nucl.\ Phys.\ B {\bf 358} (1991) 497.
 







\bibitem{Lieb}
E.~H.~Lieb, T.~Schultz and D.~Mattis,
\emph{Two soluble models of an antiferromagnetic chain,}
Annals Phys. \textbf{16} (1961), 407-466.
T.~D.~Schultz, D.~C.~Mattis and E.~H.~Lieb,
\emph{Two-dimensional Ising model as a soluble problem of many fermions,}
Rev. Mod. Phys. \textbf{36} (1964), 856-871.
B.~U.~Felderhof, 
\emph{Direct diagonalization of the transfer matrix of the zero-field free-fermion model,} 
Physica 65 (1973) 421.
{}
B.~U.~Felderhof, 
\emph{Diagonalization of the transfer matrix of the free-fermion model. II,}
Physica 66 (1973) 279. 
{}
B.~U.~Felderhof, 
\emph{Diagonalization of the transfer matrix of the free-fermion model. III,} Physica 66 (1973) 509.
    V.~V.~Bazhanov and Y.~G.~Stroganov,
  \emph{Free Fermions on Three-dimensional Lattice and Tetrahedron Equations,}
  Nucl.\ Phys.\ B {\bf 230} (1984) 435.
{}
V.~V.~Bazhanov and Y.~G.~Stroganov,
  \emph{Hidden Symmetry of Free Fermion Model. 1. Triangle Equations and Symmetric Parametrization,}
  Theor.\ Math.\ Phys.\  {\bf 62} (1985) 253
   [Teor.\ Mat.\ Fiz.\  {\bf 62} (1985) 377].
   {}
   V.~V.~Bazhanov and Y.~G.~Stroganov,
  \emph{Hidden Symmetry of the Free Fermion Model. 2. Partition Function,}
  Theor.\ Math.\ Phys.\  {\bf 63} (1985) 519
   [Teor.\ Mat.\ Fiz.\  {\bf 63} (1985) 291].
   {}
   V.~V.~Bazhanov and Y.~G.~Stroganov,
  \emph{Hidden Symmetry of the Free Fermion Model. 3. Inversion Relations,}
  Theor.\ Math.\ Phys.\  {\bf 63} (1985) 604
   [Teor.\ Mat.\ Fiz.\  {\bf 63} (1985) 417].
 R.~J.~Baxter
\emph{Free-Fermion, Checkerboard and Z-invariant Lattice Models in Statistical Mechanics,} Proc. Royal Society of London, Series A, Math. and Phys., 1986, vol. 404, nr. 1826
N.~Cramp\'e, R.~I.~Nepomechie and L.~Vinet,
\emph{Free-Fermion entanglement and orthogonal polynomials,}
J. Stat. Mech. 2019, 093101, [\arXivlink{1907.00044}].

\bibitem{MitevEtAl}
V.~Mitev, M.~Staudacher and Z.~Tsuboi,
\emph{The Tetrahedral Zamolodchikov Algebra and the ${AdS_5\times S^5}$ S-matrix,}
Commun. Math. Phys. \textbf{354} (2017) no.1, 1-30
[\arXivlink{1210.2172}].
C.~M.~Viallet,
\emph{Free Fermion Conditions and the Symmetries of Integrability},
International Journal of Modern Physics B \textbf{11} (1997), 213-221.
M. ~Wheeler,
\emph{Free fermions in classical and quantum integrable models},
[\arXivlink{1110.6703}].
P. Fendley, 
\emph{Free fermions in disguise,} 
J. Phys. A {\bf 52} (2019) 335002.
P. Fendley, 
\emph{Free parafermions,} 
J. Phys. A {\bf 47} (2014) 075001.
A.~Melikyan and G.~Weber,
\emph{The Lax pair for the fermionic Bazhanov-Stroganov $R$-operator,}
[\arXivlink{2011.03066}].
Y.~Umeno, M.~Shiroishi and M.~Wadati,
\emph{Fermionic R operator for the fermion chain model,}
J. Phys. Soc. Jap. \textbf{67} (1998), 1930.
Y.~Umeno,
\emph{Fermionic R operator and algebraic structure of 1d Hubbard model: Its application to quantum transfer matrix,}
J. Phys. Soc. Jap. \textbf{70} (2001), 2531
F.~C.~Alcaraz and R.~A.~Pimenta,
\emph{Free fermionic and parafermionic quantum spin chains with multispin interactions,}
Phys. Rev. B \textbf{102} (2020) no.12, 121101, [\arXivlink{2005.14622}]. 
F.~C.~Alcaraz and R.~A.~Pimenta,
\emph{Integrable quantum spin chains with free fermionic and parafermionic spectrum,}
 [\arXivlink{2010.01116}]. 



\bibitem{Dublin}
M.~De Leeuw, C.~Paletta, A.~Pribytok, A.~L.~Retore and A.~Torrielli,
\emph{Free Fermions, vertex Hamiltonians, and lower-dimensional AdS/CFT,}
JHEP \textbf{02} (2021), 191
[\arXivlink{2011.08217}].


\bibitem{Marius}
M.~De Leeuw, A.~Pribytok, A.~L.~Retore and P.~Ryan,
\emph{New integrable 1D models of superconductivity,}
J. Phys. A \textbf{53} (2020) no.38, 385201
[\arXivlink{1911.01439}].
M.~de Leeuw, C.~Paletta, A.~Pribytok, A.~L.~Retore and P.~Ryan,
\emph{Classifying Nearest-Neighbor Interactions and Deformations of AdS,}
Phys. Rev. Lett. \textbf{125} (2020) no.3, 031604
[\arXivlink{2003.04332}].

\bibitem{NewMarius}
M.~de Leeuw, C.~Paletta, A.~Pribytok, A.~L.~Retore and P.~Ryan,
\emph{Yang-Baxter and the Boost: splitting the difference,}
SciPost Phys. \textbf{11} (2021), 069
[\arXivlink{2010.11231}].
M.~de Leeuw, A.~Pribytok, A.~L.~Retore and P.~Ryan,
\emph{Integrable deformations of AdS/CFT,}
[\arXivlink{2109.00017}].
A.~Pribytok,
\emph{Automorphic symmetries and $ AdS_n $ integrable deformations,}
[\arXivlink{2112.10843}].
M.~Bocconcello, I.~Masuda, F.~K.~Seibold and A.~Sfondrini,
\emph{S matrix for a three-parameter integrable deformation of $AdS_3 \times S^3$ strings,}
JHEP \textbf{11} (2020), 022
[\arXivlink{2008.07603}].
J.~M.~N.~Garc\'\i{}a and L.~Wyss,
\emph{Three-parameter deformation of $\mathbbmss{R} \times S^3$ in the Landau-Lifshitz limit,}
JHEP \textbf{07} (2021), 028
[\arXivlink{2102.06419}].

\bibitem{AleSBurkhardtLePlat}
B.~Eden, D.~l.~Plat and A.~Sfondrini,\emph{Integrable bootstrap for $AdS_3/CFT_2$ correlation functions,}
JHEP \textbf{08} (2021), 049
[\arXivlink{2102.08365}].

\bibitem{AleForm}
A.~Torrielli,
\emph{A study of integrable form factors in massless relativistic $AdS_3$,}
[\arXivlink{2106.06874}].

\bibitem{Maillet}
J.~M.~Maillet and J.~Sanchez de Santos,
\emph{Drinfel'd twists and algebraic Bethe ansatz,}
[\arXivlink{q-alg/9612012}].

\bibitem{Maillet2}
N.~Kitanine, J.~M.~Maillet and V.~Terras,
\emph{Form factors of the XXZ Heisenberg spin-$\frac{1}{2}$ finite chain,}
Nucl. Phys. B \textbf{554} (1999), 647-678
[\arXivlink{math-ph/9807020}].


\bibitem{Chari:1994pz}
V.~G.~Drinfeld,
\emph{Quasi Hopf algebras,}
Alg. Anal. \textbf{1N6} (1989), 114-148.
V.~G.~Drinfeld,
\emph{Quantum groups,}
Zap. Nauchn. Semin. \textbf{155} (1986), 18-49.
V.~Chari and A.~Pressley,
\emph{A guide to quantum groups,} Cambridge University Press 1994.


\bibitem{Drinfeld:1987sy}
V.~G.~Drinfeld,
\emph{A New realization of Yangians and quantized affine algebras,}
Sov. Math. Dokl. \textbf{36} (1988), 212-216

\bibitem{Khoroshkin:1994uk}
S.~M.~Khoroshkin and V.~N.~Tolstoi,
\emph{Yangian double,}
Lett. Math. Phys. {\bf 36} (1996) 373
[\arXivlink{hep-th/9406194}].

\bibitem{Koro2}
J.~Cai, S.~Wang, K.~Wu and C.~Xiong, \emph{Universal-matrix of the Super Yangian Double $DY(\alg{gl}(1| 1))$,} Commun. Theo. Phys. {\bf 29} (1998) 173-176.
S.~Moriyama and A.~Torrielli,
\emph{A Yangian double for the AdS/CFT classical r-matrix,}
JHEP \textbf{06} (2007), 083
[\arXivlink{0706.0884}].

\bibitem{unive0}
N.~Beisert and F.~Spill,
\emph{The Classical r-matrix of AdS/CFT and its Lie Bialgebra Structure,}
Commun. Math. Phys. \textbf{285} (2009), 537-565
[\arXivlink{0708.1762}].

\bibitem{unive}
G.~Arutyunov, M.~de Leeuw and A.~Torrielli,
\emph{Universal blocks of the AdS/CFT Scattering Matrix,}
JHEP \textbf{05} (2009), 086
[\arXivlink{0903.1833}].

\bibitem{korotwi}
S.~M.~Khoroshkin and V.~N.~Tolstoi,
\emph{Twisting of quantum (super)algebras: Connection of Drinfeld's and Cartan-Weyl realizations for quantum affine algebras,}
[\arXivlink{hep-th/9404036}].

\bibitem{secret}
T.~Matsumoto, S.~Moriyama, and A.~Torrielli,
{\em {A secret symmetry of the AdS/CFT $S$-matrix},}
{JHEP {\bf 09} (2007)
  099} [\arXivlink{0708.1285}].
 \bibitem{segreto}
  M.~de Leeuw, T.~Matsumoto, S.~Moriyama, V.~Regelskis and A.~Torrielli,
  {\em Secret Symmetries in AdS/CFT,}
  Phys.\ Scripta {\bf 02} (2012) 028502
  [\arXivlink{1204.2366}].

\bibitem{bgood}
N.~Gromov, F.~Levkovich-Maslyuk and G.~Sizov,
\emph{New Construction of Eigenstates and Separation of Variables for SU(N) Quantum Spin Chains,}
JHEP \textbf{09} (2017), 111
[\arXivlink{1610.08032}].
{}
N.~Gromov, F.~Levkovich-Maslyuk, P.~Ryan and D.~Volin,
\emph{Dual Separated Variables and Scalar Products,}
Phys. Lett. B \textbf{806} (2020), 135494
[\arXivlink{1910.13442}].

 \bibitem{fofa}
S.~Britton and S.~Frolov,
  {\em Free field representation and form factors of the chiral Gross-Neveu model,}
  JHEP {\bf 1311} (2013), 076
  [\arXivlink{1305.6252}].
  T.~Klose and T.~McLoughlin,
  {\em Worldsheet Form Factors in AdS/CFT,}
  Phys.\ Rev.\ D {\bf 87} (2013),  026004
  [\arXivlink{1208.2020}]. 
  T.~Klose and T.~McLoughlin,
  {\em Comments on World-Sheet Form Factors in AdS/CFT,}
  J.\ Phys.\ A {\bf 47} (2014),  055401
  [\arXivlink{1307.3506}].
  
\bibitem{AlerevY}
A.~Torrielli,
\emph{Yangians, S-matrices and AdS/CFT,}
J. Phys. A \textbf{44} (2011), 263001
[\arXivlink{1104.2474}].

\bibitem{les}
M.~De Leeuw, B.~Eden and A.~Sfondrini,
\emph{Bound state scattering simplified,}
Phys. Rev. D \textbf{102} (2020) no.12, 126001
[\arXivlink{2008.01378}].

\bibitem{twists}
E.~Pomoni, R.~Rabe and K.~Zoubos,
\emph{Dynamical spin chains in 4D $ \mathcal{N}= 2$ SCFTs,}
JHEP \textbf{08} (2021), 127
[\arXivlink{2106.08449}].
\bibitem{vanTongeren:2021jhh}
S.~J.~van Tongeren and Y.~Zimmermann,
\emph{Do Drinfeld twists of $AdS_5 \times S^5$ survive light-cone quantization?,}
[\arXivlink{2112.10279}].
R.~Borsato, S.~Driezen and J.~L.~Miramontes,
\emph{Homogeneous Yang-Baxter deformations as undeformed yet twisted models,}
[\arXivlink{2112.12025}].

  \end{thebibliography}
\end{document}